# Higgs Boson Searches at the LHC Beyond the Standard Model

## André Sopczak
### on behalf of the ATLAS and CMS Collaborations

Institute of Experimental and Applied Physics,
Czech Technical University in Prague, Czech Republic

## Abstract

The latest results of Higgs boson searches beyond the Standard Model from the ATLAS and CMS experiments are reviewed. This includes searches for additional neutral, charged, and double charged Higgs-like bosons, searches for dark matter produced in association with a Higgs boson, and searches for new physics in Higgs boson production and decay processes. Interpretations are given within the hMSSM, a special parameterization of the Minimal Supersymmetric extension of the Standard Model, in which the mass of the lightest Higgs boson is set to the value of 125 GeV measured at the LHC.

*Review*

# Higgs Boson Searches at the LHC Beyond the Standard Model

## André Sopczak

Institute of Experimental and Applied Physics, Czech Technical University in Prague, Husova 240/5, 11000 Prague 1, Czech Republic; andre.sopczak@cern.ch

**Abstract:** The latest results of Higgs boson searches beyond the Standard Model from the ATLAS and CMS experiments are reviewed. This includes searches for additional neutral, charged, and double charged Higgs-like bosons, searches for dark matter produced in association with a Higgs boson, and searches for new physics in Higgs boson production and decay processes. Interpretations are given within the hMSSM, a special parameterization of the Minimal Supersymmetric extension of the Standard Model, in which the mass of the lightest Higgs boson is set to the value of 125 GeV measured at the LHC.

## Contents

## 1. Introduction

Since the discovery of the Higgs boson with the properties of the Standard Model (SM) in 2012, many searches for Higgs bosons beyond the Standard Model have been made. These searches can be organized according to different proposed extensions of the Higgs sector [1]. Two-Higgs-Doublet Models (2HDMs) introduce five Higgs bosons: two scalar ($h, H$) neutral, a pseudoscalar ($A$) neutral, and two ($H^{\pm}$) charged particles. The Minimal Supersymmetric extension of the SM (MSSM) is a 2HDM that can contain additional particles within reach of the Large Hadron Collider (LHC), namely the superpartners of SM particles. The 2HDMs with an extra light singlet ($a$) bring new phenomena through the opening of the decay $H \to aa$. Three-HDMs (3HDMs) enlarge the Higgs boson sector, having three charge-parity $CP$-even and two $CP$-odd neutral Higgs bosons, as well as two charged Higgs bosons. In models with an additional U(1)$_D$ gauge symmetry, a pseudoscalar ($A$) and charged Higgs bosons ($H^{\pm}$) are predicted. In Type-II seesaw models, left–right symmetric models (LRSMs), the Zee–Babu neutrino mass model, 3-3-1 models, and the Georgi–Machacek model, doubly charged Higgs ($H^{++}$) are predicted.

A previous review on the SM Higgs boson research was given at the 2019 International Conference on Precision Physics and Fundamental Physical Constants, FFK2019 [2].

This review is structured as follows. Section 2 describes Higgs boson decays, Section 3 is devoted to the enhanced production modes, Section 4 describes Higgs boson resonance production searches, Section 5 reports on the additional neutral Higgs boson production, Section 6 reports on the Higgs boson dark matter production, and Section 7 reports on the charged Higgs boson production searches. Interpretations and an outlook are given in Section 8.

### 1.1. LHC, Experiments, and data taking

The data from proton–proton collisions at the LHC at CERN have been recorded with the ATLAS and CMS experiments in three data taking periods:

- LHC Run-1: 2010–2012, discovery of the Higgs boson [3,4];
- LHC Run-2: 2015–2018, results presented here;
- LHC Run-3: 2022–2025, ongoing data taking.

Figure 1 illustrates the layout of the LHC (Figure 1, left) and of the ATLAS and CMS experiments (Figure 1, right). The LHC location of the underground area is shown, as well as the terrain.

Figure 2, left, shows the luminosity delivered to the ATLAS experiment from proton–proton collisions [5], while Figure 2, right, shows a schematic view of particle identification in the ATLAS or CMS detectors.

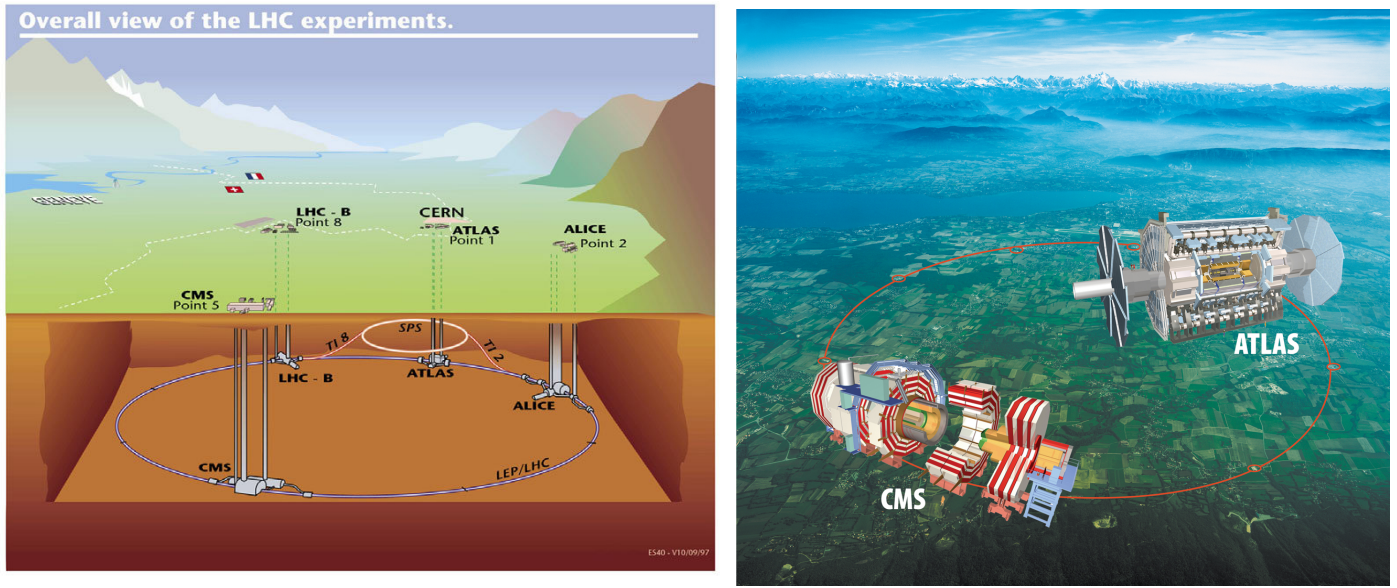

**Figure 1. Left**: LHC underground and terrain areas with four experiments. **Right**: ATLAS and CMS experiments.

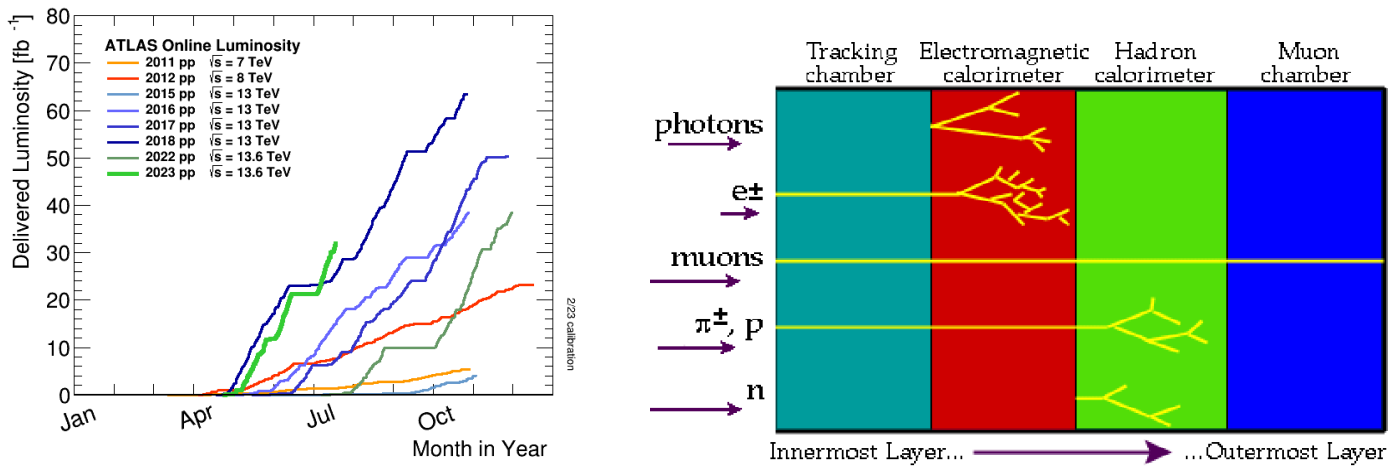

**Figure 2. Left**: Delivered luminosity at the ATLAS experiment [5] (CC BY 4.0). **Right**: Particle identification schema by different subsystems.

Event displays of typical recorded proton-proton collisions are shown in Figure 3 [6,7]. The proton–proton collisions in the center of the detectors lead to different reactions, including the production of the SM Higgs boson and potentially the production of the Higgs boson predicted in extended models.

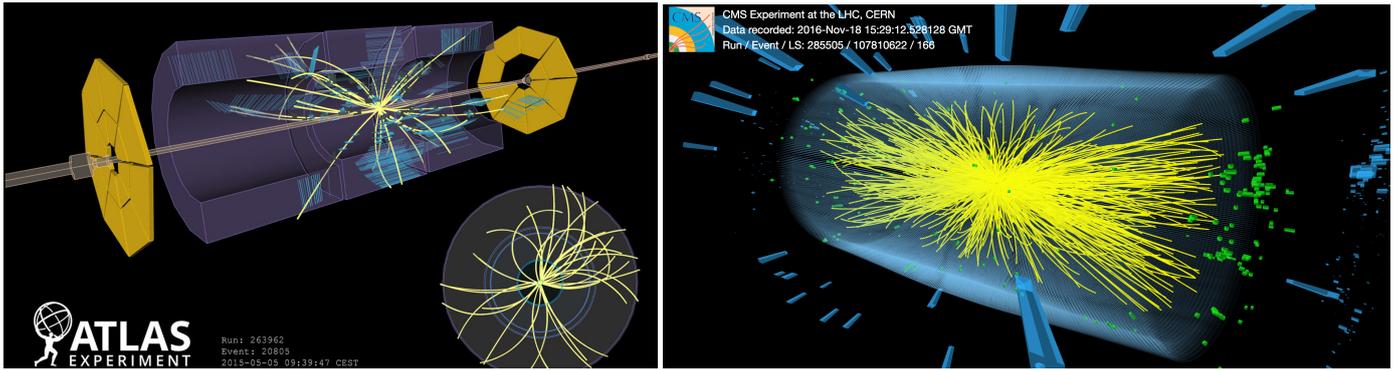

**Figure 3.** Event displays from (**left**) ATLAS [6] (CC BY 4.0) and (**right**) CMS [7] (CC BY 4.0) experiments.

*1.2. Production Modes*

The analyses presented here are mostly based on the full LHC Run-2 dataset. The evolution of the integrated luminosity recorded by the CMS experiment as a function of time is shown in Figure 4, left, along with the number of expected SM Higgs boson events in the production modes $ggH$, $VBF$, $VH$, $ttH$, $tH$, and $HH$ (Figure 4, right), where $g$ denotes gluon, $VBF$ denotes vector boson fusion, and $t$ denotes the top quark. The corresponding Feynman diagrams are shown in Figures 5–7 along with the production cross-sections for 13 TeV center-of-mass energy [8] and the resulting expected rates for 140 fb$^{-1}$ integrated luminosity [9]. The expected rates can vary in different extensions of the SM.

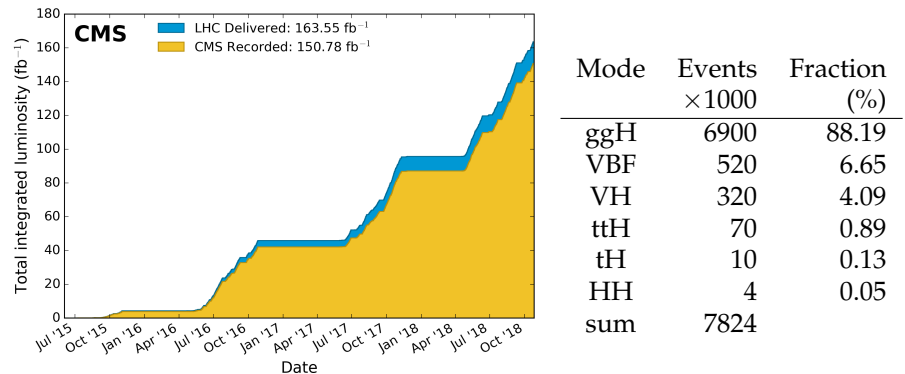

| Mode | Events ×1000 | Fraction (%) |
|------|------|------|
| ggH | 6900 | 88.19 |
| VBF | 520 | 6.65 |
| VH | 320 | 4.09 |
| ttH | 70 | 0.89 |
| tH | 10 | 0.13 |
| HH | 4 | 0.05 |
| sum | 7824 | |

**Figure 4. Left**: LHC Run-2 integrated luminosity [10] (CC BY 4.0). **Right**: Expected numbers of Higgs boson events for 140 fb$^{-1}$ integrated luminosity per experiment.

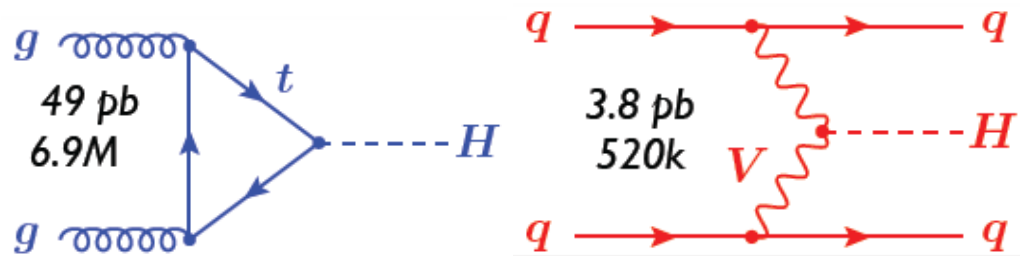

**Figure 5.** Feynman diagrams for $ggH$ (**left**) and $VBF$ (**right**) Higgs boson production modes (see text for details). The production cross-sections (in pb) are given for 13 TeV center-of-mass energy [8] and the expected rates ("M" stands for ×10$^6$ and "k" stands for ×10$^3$) are given for 140 fb$^{-1}$ integrated luminosity [9].

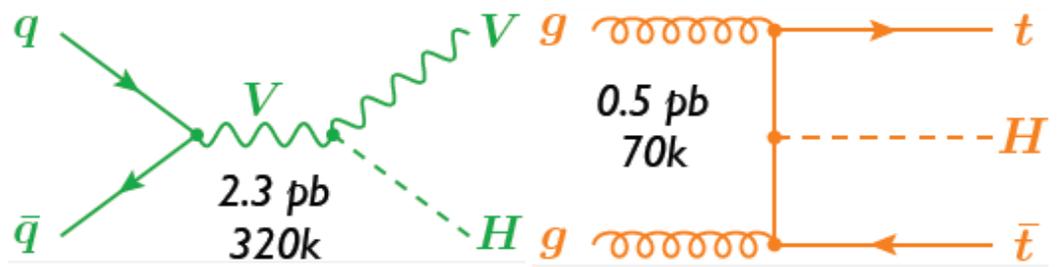

**Figure 6.** Feynman diagrams for *VH* (**left**) and *ttH* (**right**) Higgs boson production modes (see text for details). The production cross-sections (in pb) are given for 13 TeV center-of-mass energy [8] and the expected rates ("k" stands for $\times 10^3$) are given for 140 fb$^{-1}$ integrated luminosity [9].

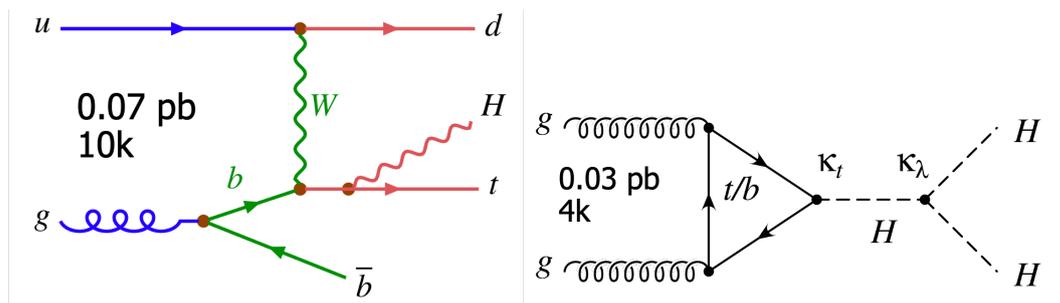

**Figure 7.** Feynman diagrams for *tH* (**left**) and *HH* (**right**) Higgs boson production modes (see text for details). The production cross-sections (in pb) are given for 13 TeV center-of-mass energy [8] and the expected rates ("k" stands for $\times 10^3$) are given for 140 fb$^{-1}$ integrated luminosity [9].

### 1.3. Decay Branching Ratios

SM predictions for the Higgs boson branching ratios and precision measurements for several Higgs boson decay modes are summarized in Figure 8 [8,11]. Physics beyond the SM (BSM) may change these ratios. After establishing the main decay channels and striving to precisely characterize them, the experimental focus is now on the measurements of the rare decay modes that can be enhanced in extended models.

| Decay channel | BR (%) |
|---|---|
| bb | $57.63 \pm 0.70$ |
| WW | $22.00 \pm 0.33$ |
| gg | $8.15 \pm 0.42$ |
| $\tau\tau$ | $6.21 \pm 0.09$ |
| cc | $2.86 \pm 0.09$ |
| ZZ | $2.71 \pm 0.04$ |
| $\gamma\gamma$ | $0.227 \pm 0.005$ |
| $Z\gamma$ | $0.157 \pm 0.009$ |
| ss | $0.025 \pm 0.001$ |
| $\mu\mu$ | $0.0216 \pm 0.0004$ |

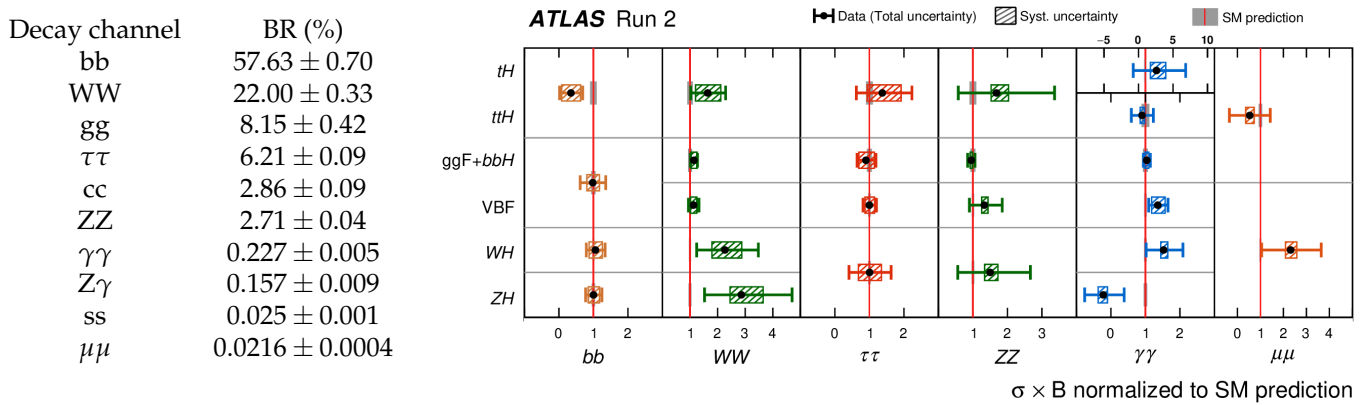

**Figure 8. Left**: Theoretical predictions of Higgs boson decay branching ratios in the SM [8]. **Right**: Cross-section times branching ratio measurements [11] (CC BY 4.0).

### 1.4. Relation of Coupling to Fermion/Vector Boson Mass

The relationships between couplings and fermions and vector bosons as a function of the particle masses measured by ATLAS and CMS experiments [11,12] are shown in Figure 9. The SM expectation and the measurements are in remarkable agreement within the uncertainties (the smallest of which is about 3% for *WW/ZZ* boson couplings).

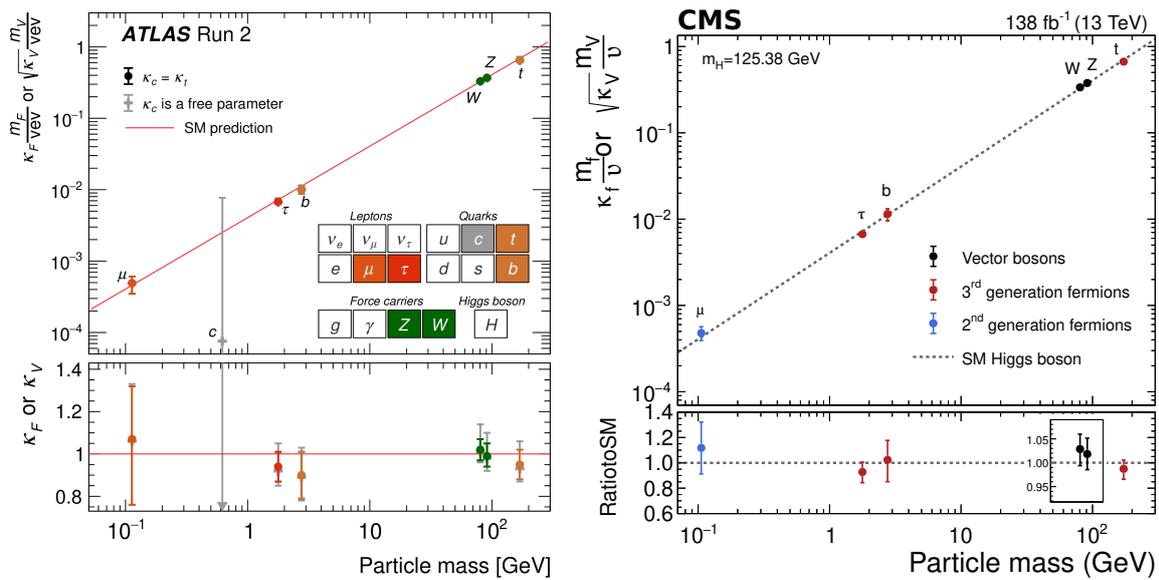

**Figure 9.** Relationships between couplings and fermion/vector boson masses by (**left**) ATLAS [11] (CC BY 4.0) and (**right**) CMS [12] (CC BY 4.0) experiments.

### 1.5. Higgs Boson Width

The width of the Higgs boson can be measured using the $ZZ$ production mode and subsequent decay, including virtual Higgs boson contributions and its interference [13]. The two sensitive processes are $ZZ \to 4\ell$ and $ZZ \to 2\ell 2\nu$, with $\ell = e$ or $\mu$. Figure 10 shows Feynman diagrams of $ZZ$ ($V = Z$) production and the simulated $ZZ$ differential cross-section as a function of the four-lepton invariant mass [13]. The contribution of the Higgs boson allows determing the Higgs boson width. The ATLAS experiment indirect limit on the width measurement is $\Gamma = 3.4^{+3.3}_{-2.5}$ MeV, with an observed upper limit of 10.5 MeV, for an expected value of 10.9 MeV at 95% CL [14] and $\Gamma = 3.2^{+2.4}_{-1.7}$ MeV for the CMS [13] experiment. Both measurements are in good agreement with the SM prediction $\Gamma = 4.1$ MeV [8]. The corresponding likelihood is shown in Figure 11 [13,14], assuming no new particles enter in the production of the virtual Higgs boson.

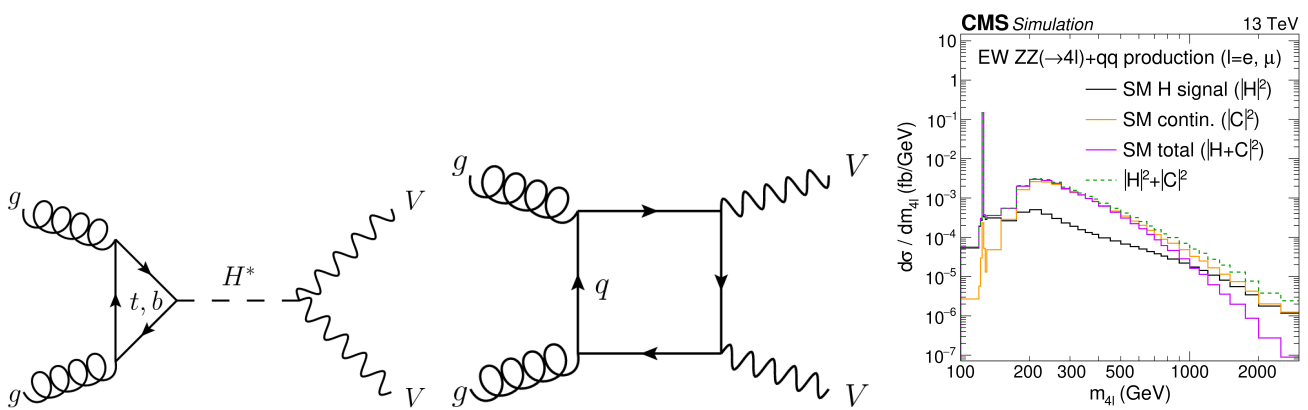

**Figure 10. Left** and **Middle**: Feynman diagrams of $VV$ production. **Right**: Simulated differential $ZZ$ production cross-section as a function of the four-lepton invariant mass [13] (CC BY 4.0).

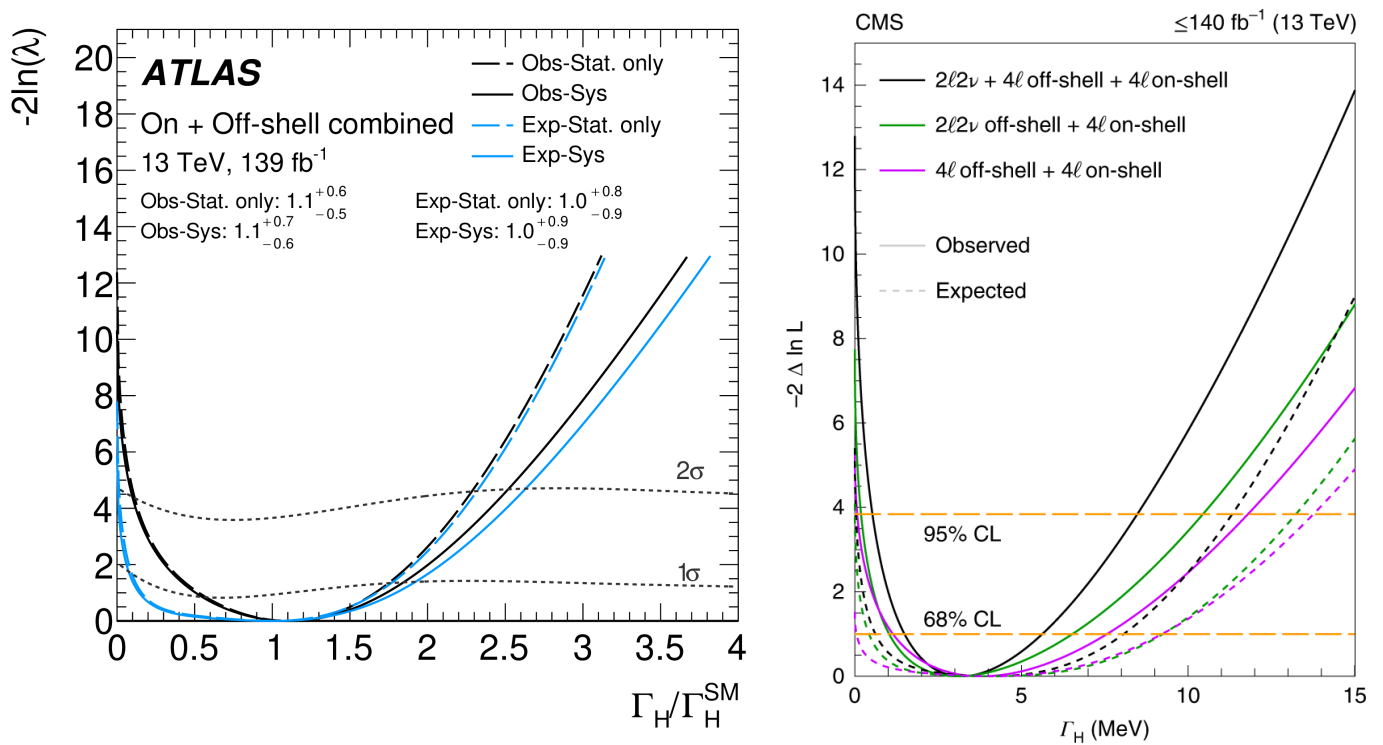

**Figure 11.** Likelihood for Higgs boson width measurements by (**left**) ATLAS [14] (CC BY 4.0) and (**right**) CMS [13] (CC BY 4.0) experiments.

### 1.6. Higgs Boson CP Properties

A study of the *CP* properties of the interaction between the Higgs boson and tau-leptons (τs) was performed, based on the reconstruction of the planes of tau decay products. The results of the observed signal strength $\mu_{\tau\tau}$ versus the *CP*-mixing angle $\phi_\tau$ are shown in Figure 12, right [15]. No indication of a non-SM contribution is observed. To clarify, Figure 12, left, presents the definition of the $\varphi_{CP}^*$ angle and Figure 12, middle, shows the data and simulation distributions. The Lagrangian defining $\phi_\tau$ and $\kappa_\tau$ reads $\mathcal{L}_{H\tau\tau} = -\frac{m_\tau}{v}\kappa_\tau(\cos\phi_\tau \bar{\tau}\tau + \sin\phi_\tau \bar{\tau} i \gamma_5 \tau)H$ [15], and $\mu_{\tau\tau}$ is the ratio of the signal strength to the SM expectation.

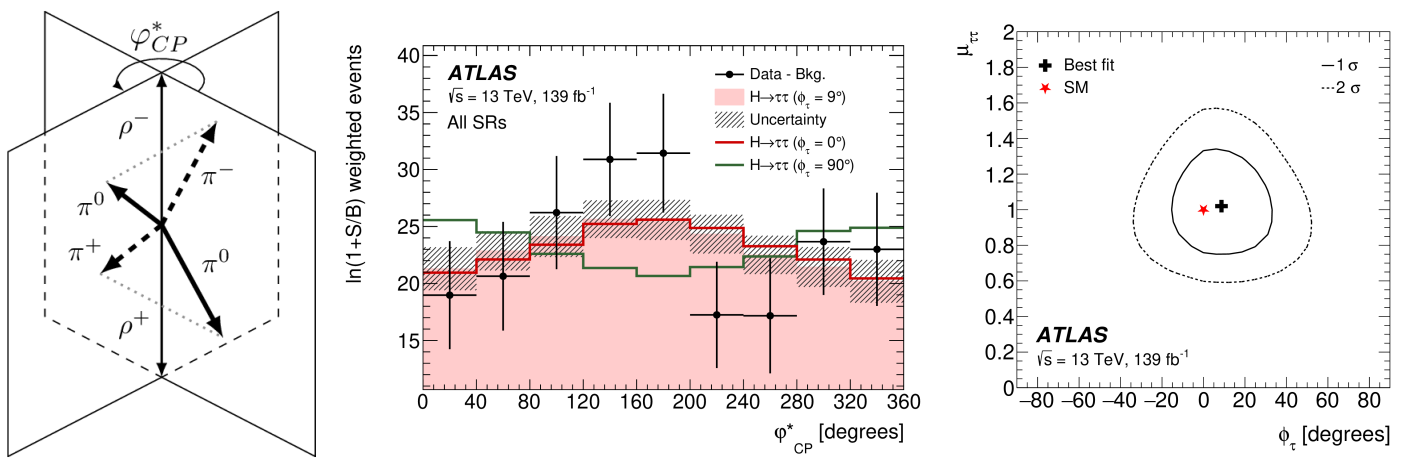

**Figure 12.** **Left**: Definition of the angle between τ-decay products. **Middle**: Measured and simulated angle distributions between τ-decay products [15] (CC BY 4.0). **Right**: Observed and expected signal strength, $\mu_{\tau\tau}$, versus the *CP*-mixing angle $\phi_\tau$ [15] (CC BY 4.0).

The *CP* structure of the Higgs boson top quark Yukawa coupling was also measured in $tH$ and $ttH$ production [16]. The parameters $\kappa_t = 0$ (*CP*-even) and $\kappa_t = 1$ (*CP*-odd) are used, and other parameters are fixed to their SM values. Figure 13, left, compares the distributions of the $ttH$ invariant mass for the pure *CP*-even and *CP*-odd assumptions and Figure 13, middle, shows the results of the measured $\kappa_t$ and $\tilde{\kappa}_t$ values [16].

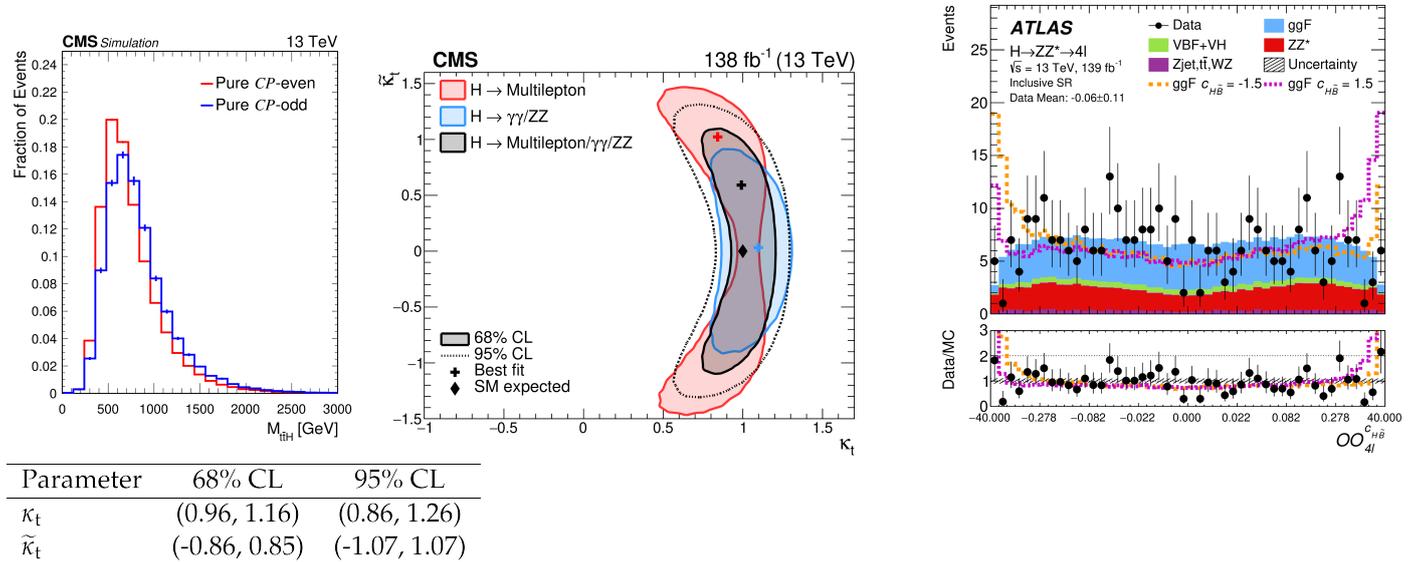

| Parameter | 68% CL | 95% CL |
|---|---|---|
| $\kappa_t$ | (0.96, 1.16) | (0.86, 1.26) |
| $\tilde{\kappa}_t$ | (-0.86, 0.85) | (-1.07, 1.07) |

**Figure 13. Left**: Comparison of the distributions of the $ttH$ invariant mass for the pure *CP*-even and *CP*-odd assumptions [16] (CC BY 4.0). **Middle**: Observed and SM expected $\kappa_t$ and $\tilde{\kappa}_t$ values [16] (CC BY 4.0). **Right**: Decay level observable for *CP*-invariance in $ggF$ $VBF$ production for $H \to ZZ \to 4\ell$ [17] (CC BY 4.0). **Lower**: Limits on measured $\kappa_t$ and $\tilde{\kappa}_t$ values [16]. See text for details.

The *CP* properties in $ggF$ and $VBF$ production for $H \to ZZ \to 4\ell$ have been studied in Ref. [17]. Matrix element-based optimal observables were used to constrain *CP*-odd couplings beyond the SM in the framework of SM effective field theory [18]. The tested *CP*-invariance of Higgs boson interactions with $VH$ ($V = W, Z$) were performed. Results are compatible with the SM expectation, and no hint of BSM physics was observed. Figure 13, right, compares data and simulation for $H \to ZZ \to 4\ell$ [17].

## 2. Higgs Boson Decay

Higgs boson couplings beyond the already observed $tt$, $ZZ$, $WW$, $bb$, $\tau\tau$, and $\mu\mu$ couplings are discussed here.

### 2.1. $H \to ee$

Higgs boson decays into a pair of electrons having a quite small branching fraction in the SM BR($H \to ee$) $\approx 5 \times 10^{-9}$. The branching fraction could be enhanced in the 2HDM, and experimental limits are set to BR($H \to ee$) $< 3.0 \times 10^{-4}$ ($3.0 \times 10^{-4}$ expectation) at 95% CL [19] when the various measurements are combined, as shown in Figure 14, left. The reconstructed di-electron mass distribution does not yet show a visible signal due to the expected low rate; see Figure 14, right.

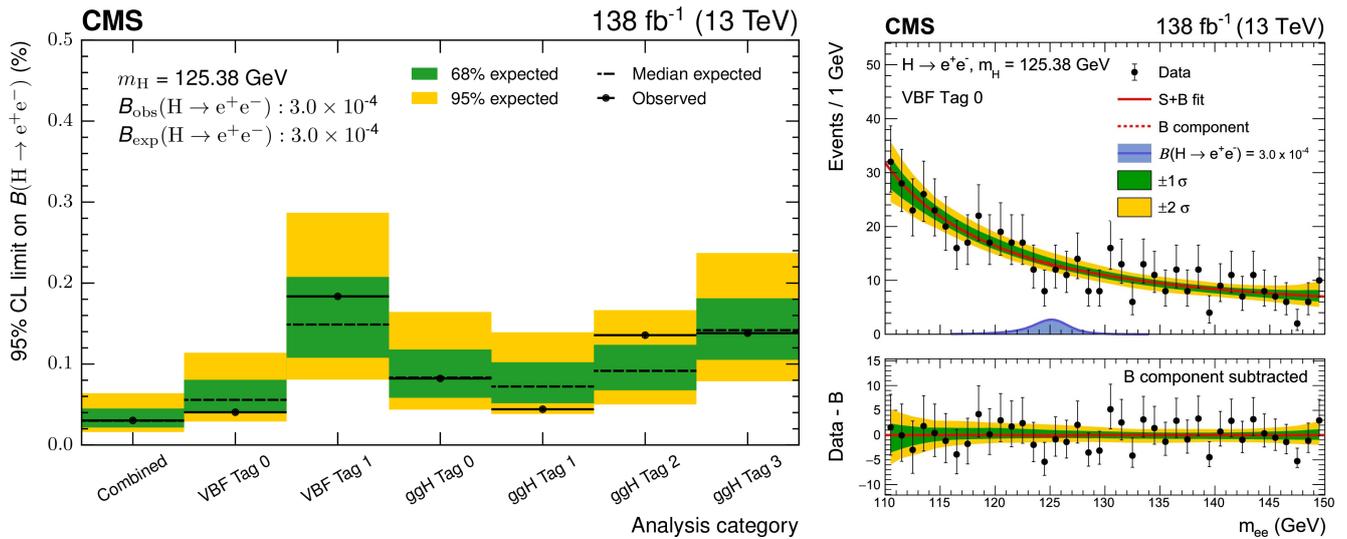

**Figure 14. Left**: Upper limits on the $H \to ee$ branching ratio in various search channels and those combined [19] (CC BY 4.0). **Right**: Reconstructed di-electron mass distributon [19] (CC BY 4.0).

### 2.2. $H \to Z\gamma$

The $H \to Z\gamma$ SM Higgs boson decay modes are within experimental reach. The ATLAS Collaboration obtains a significance of 2.2 (1.2 exp.) standard deviation (s.d.), and the best-fit value for the signal yield normalized to the SM prediction is $\mu = 2.0^{+1.0}_{-0.9}$ [20], while the CMS Collaboration obtains a significance of 2.7 (1.2 exp.) s.d. and $\mu = 2.4 \pm 0.9$ [21], where the statistical component of the uncertainty is dominant. Figure 15 shows the $Z\gamma$ distribution from ATLAS [20] (Figure 15, left) and CMS [21] (Figure 15, right) experiments, as well as an event display of an ATLAS experiment candidate event (Figure 15, middle). A combination of the data from the ATLAS and CMS Collaborations leads to a 3.4 (1.6 exp.) s.d. evidence with $\mu = 2.2 \pm 0.7$ [22] (Figure 16, left).

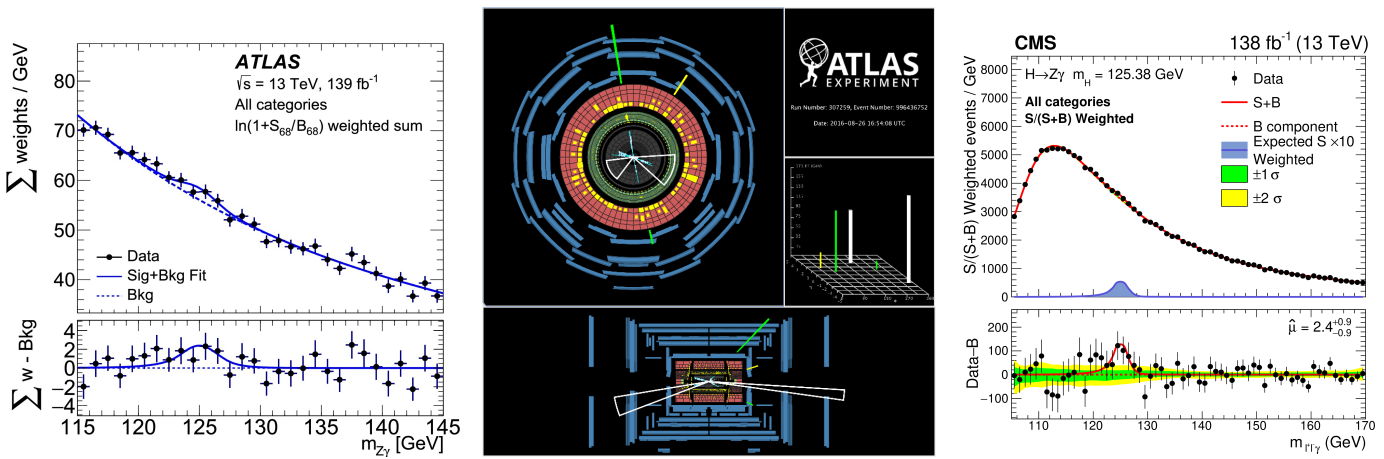

**Figure 15. Left**: ATLAS experiment $Z\gamma$ invariant-mass distribution [20] (CC BY 4.0). **Middle**: ATLAS experiment $Z\gamma$ ($Z \to ee$) candidate event display [20] (CC BY 4.0). **Right**: CMS experiment $Z\gamma$ invariant-mass distribution [21] (CC BY 4.0).

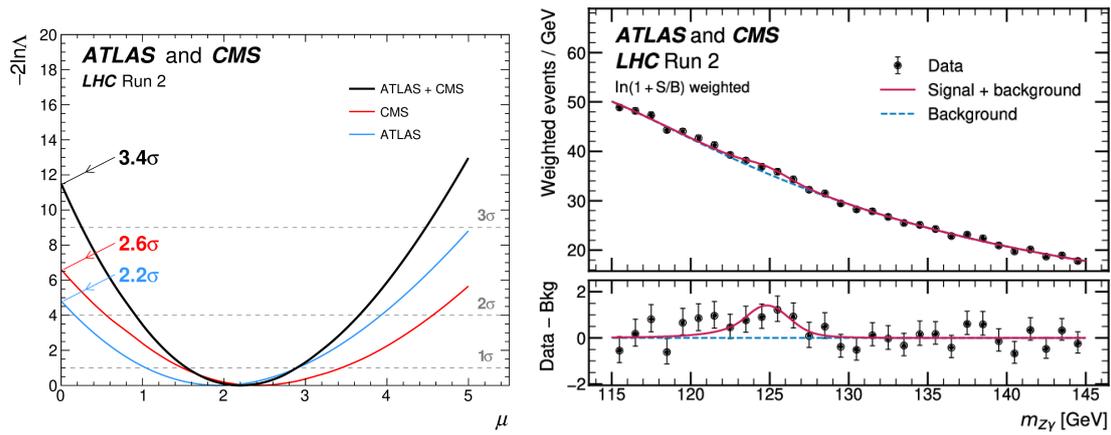

**Figure 16. Left**: Negative profile log-likelihood test statistic as a function of the $Z\gamma$ signal strength $\mu$ [22] (CC BY 4.0),where $\sigma$ denotes standard deviation. **Right**: ATLAS and CMS experiments $Z\gamma$ invariant mass distribution [22] (CC BY 4.0).

### 2.3. $H \to J/\psi J/\psi,\ \Upsilon\Upsilon,\ ZJ/\psi$

Further rare Higgs boson decays lead to four-muon final states. Experimentally, this has a clean signature owing to considerably small SM background (Figure 17). The expected sensitivity for the branching fractions is several orders of magnitude below the current experimental sensitivity. The observation of such decays would indicate physics beyond the SM. Observed and expected limits at 95% CL are in good agreement (Figure 17, upper) [23].

| Process | Observed | Expected | Observed | |
|---|---|---|---|---|
| Higgs boson channel | Longitudinal | Longitudinal | Unpolarized | Transverse |
| $\mathcal{B}(H \to ZJ/\psi)$ | $1.9 \times 10^{-3}$ | $(2.6^{+1.1}_{-0.7}) \times 10^{-3}$ | $2.4 \times 10^{-3}$ | $2.8 \times 10^{-3}$ |
| $\mathcal{B}(H \to Z\psi(2S))$ | $6.6 \times 10^{-3}$ | $(7.1^{+2.8}_{-2.0}) \times 10^{-3}$ | $8.3 \times 10^{-3}$ | $9.4 \times 10^{-3}$ |
| $\mathcal{B}(H \to J/\psi J/\psi)$ | $3.8 \times 10^{-4}$ | $(4.6^{+2.0}_{-0.4}) \times 10^{-4}$ | $4.7 \times 10^{-4}$ | $5.2 \times 10^{-4}$ |
| $\mathcal{B}(H \to \psi(2S)J/\psi)$ | $2.1 \times 10^{-3}$ | $(1.4^{+0.6}_{-0.4}) \times 10^{-3}$ | $2.6 \times 10^{-3}$ | $2.9 \times 10^{-3}$ |
| $\mathcal{B}(H \to \psi(2S)\psi(2S))$ | $3.0 \times 10^{-3}$ | $(3.3^{+1.5}_{-0.9}) \times 10^{-3}$ | $3.6 \times 10^{-3}$ | $4.7 \times 10^{-3}$ |
| $\mathcal{B}(H \to \Upsilon(nS)\Upsilon(mS))$ | $3.5 \times 10^{-4}$ | $(3.6^{+0.2}_{-0.3}) \times 10^{-4}$ | $4.3 \times 10^{-4}$ | $4.6 \times 10^{-4}$ |
| $\mathcal{B}(H \to \Upsilon(1S)\Upsilon(1S))$ | $1.7 \times 10^{-3}$ | $(1.7^{+0.1}_{-0.1}) \times 10^{-3}$ | $2.0 \times 10^{-3}$ | $2.2 \times 10^{-3}$ |
| Z boson channel | | | | |
| $\mathcal{B}(Z \to J/\psi J/\psi)$ | $11 \times 10^{-7}$ | $(9.5^{+3.8}_{-2.6}) \times 10^{-7}$ | $14 \times 10^{-7}$ | $16 \times 10^{-7}$ |
| $\mathcal{B}(Z \to \Upsilon(nS)\Upsilon(mS))$ | $3.9 \times 10^{-7}$ | $(4.0^{+0.3}_{-0.3}) \times 10^{-7}$ | $4.9 \times 10^{-7}$ | $5.6 \times 10^{-7}$ |
| $\mathcal{B}(Z \to \Upsilon(1S)\Upsilon(1S))$ | $1.8 \times 10^{-6}$ | $(1.8^{+0.1}_{-0.0}) \times 10^{-6}$ | $2.2 \times 10^{-6}$ | $2.4 \times 10^{-6}$ |

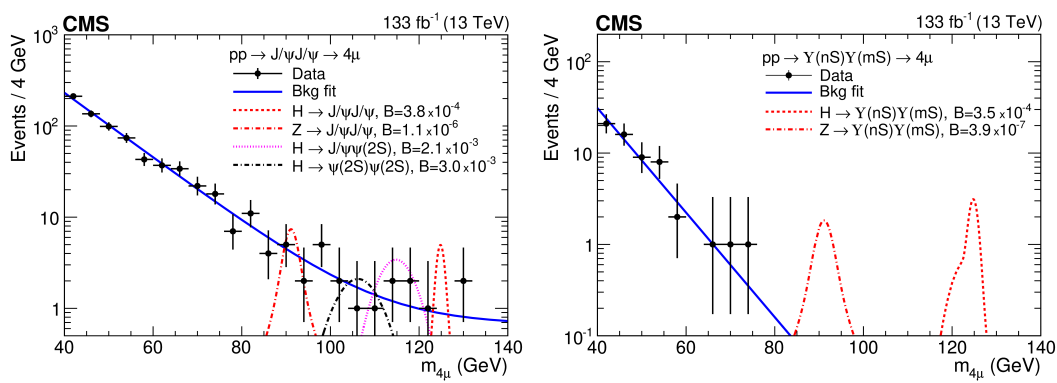

**Figure 17. Upper**: Observed and expected limits on Higgs boson branching ratios. **Lower left**: Four-muon invariant mass for $H \to J/\psi J/\psi$ searches. The limits are given for different assumptions on the polarization of the intermediate particles. The signal distributions are shown for branching ratios at the observed limit [23] (CC BY 4.0). **Lower right**: Four-muon invariant mass for $H \to \Upsilon\Upsilon$ searches [23] (CC BY 4.0).

### 2.4. $H \to \omega\gamma$, $K^*\gamma$

Searches for Higgs boson decay to $\omega\gamma$, $K^*\gamma$ probe flavor-conserving and flavor-violating Higgs boson couplings to light quarks [24,25]. Figure 18, upper left and middle, shows relevant Feynman diagrams and Figure 18, lower, shows invariant mass distributions, observed and expected limits. No excess of branching ratio is observed (see Figure 18, upper right).

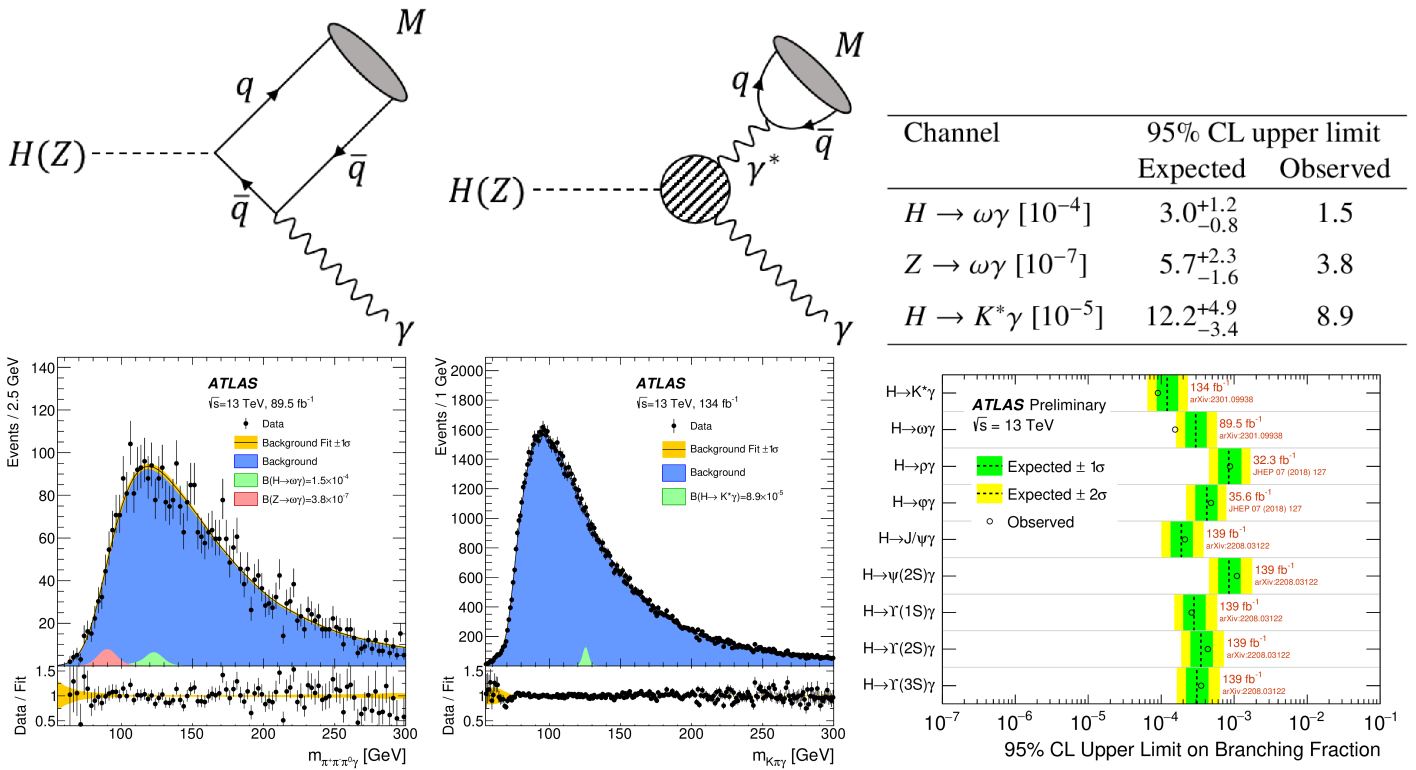

| Channel | 95% CL upper limit | |
| --- | --- | --- |
| | Expected | Observed |
| $H \to \omega\gamma$ [$10^{-4}$] | $3.0^{+1.2}_{-0.8}$ | 1.5 |
| $Z \to \omega\gamma$ [$10^{-7}$] | $5.7^{+2.3}_{-1.6}$ | 3.8 |
| $H \to K^*\gamma$ [$10^{-5}$] | $12.2^{+4.9}_{-3.4}$ | 8.9 |

**Figure 18. Upper left** and **middle**: Feynman diagrams for the associated production of a meson and a photon in Higgs boson decays [24]. **Upper right**: Observed and expected limits: $H \to \omega\gamma$, $K^*\gamma$ [24]. **Lower**: Invariant masses of mesons and photons (**left** and **middle**) [24] (CC BY 4.0) and a summary of the limits (**right**) [25] (CC BY 4.0).

### 2.5. Lepton Flavor Violation (LFV)

Searches for LFV processes $H \to e\tau$, $\mu\tau$ led to limits on the Higgs boson branching fractions BR($H \to e\tau$) < 0.20% (0.12% exp.) and BR($H \to \mu\tau$) < 0.18% (0.09% exp.) at 95% CL. No measurement is compatible with LFV. Figure 19, upper left, shows an observed invariant-mass distribution along with the simulated BR($H \to e\tau$) and BR($H \to \mu\tau$) signals, which were observed and, in the SM, expected to be BR($H \to e\tau$) and BR($H \to \mu\tau$) results (Figure 19, upper right). Figure 19, lower, shows limits on these branching ratios set at 95% CL [26]. CMS experiment searches for LFV processes $H \to e\mu$ led to branching ratio limits 4.4 (4.7 exp.) $\times 10^{-5}$ at 95% CL [27]. An excess of events over the expected background is observed at an electron-muon invariant mass of approximately 146 GeV with a global (local) significance of 2.8 (3.8) s.d. Figure 20, left, shows the $e\mu$-invariant-mass distribution and Figure 20, middle, shows the resulting limits on the Higgs boson branching fractions ($H \to e\mu$) at 95% CL [27]. The ATLAS Collaboration observed no significant signal, in agreement with the SM expectation (Figure 20, right) [28].

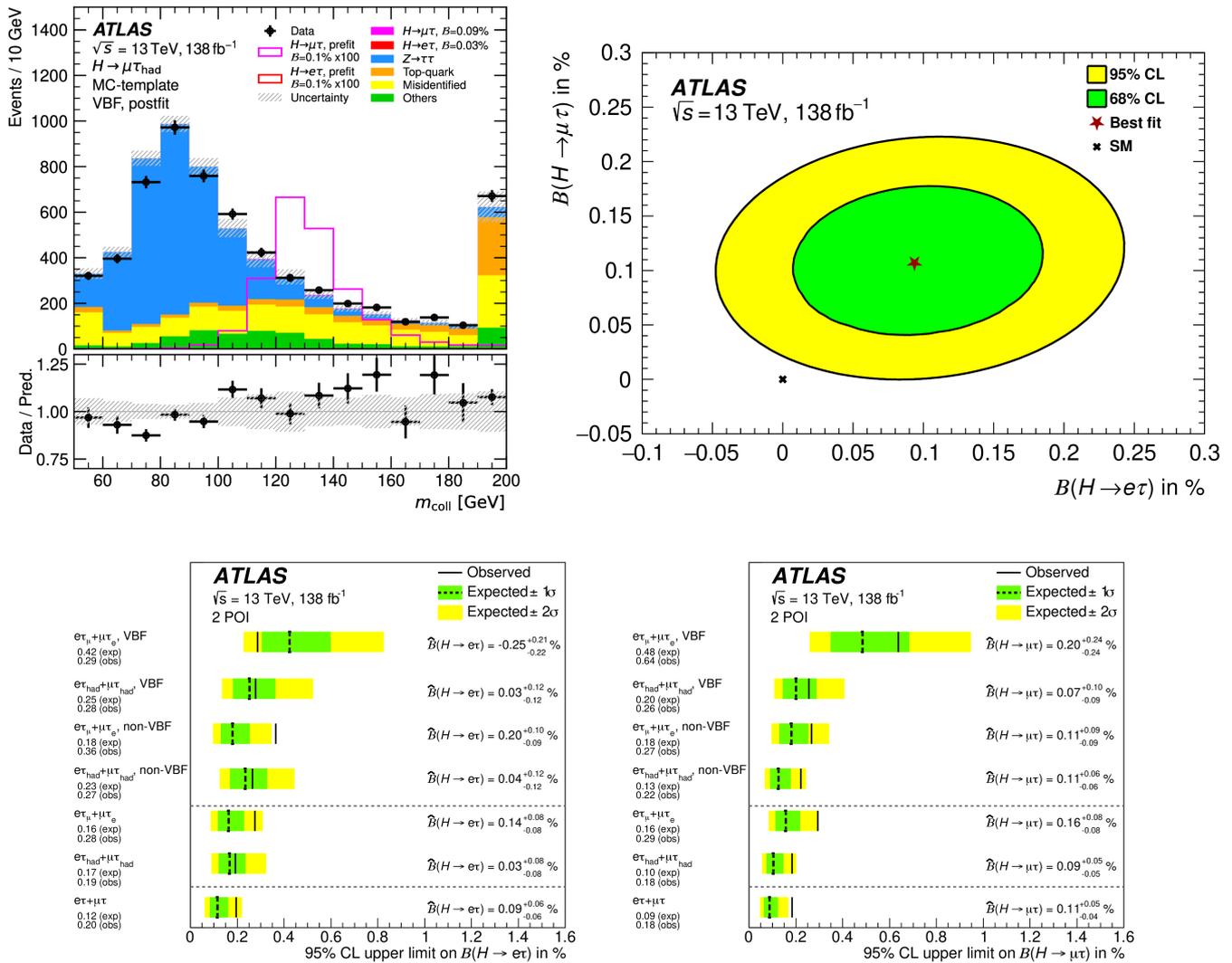

**Figure 19.** **Upper left**: Invariant-mass distribution along with simulated BR($H \rightarrow e\tau$) and BR($H \rightarrow \mu\tau$) signals [26] (CC BY 4.0). **Upper right**: Observed results and SM expectations for BR($H \rightarrow e\tau$) and BR($H \rightarrow \mu\tau$) [26] (CC BY 4.0). **Lower**: Limits on BR($H \rightarrow e\tau$) and BR($H \rightarrow \mu\tau$) at 95% CL [26] (CC BY 4.0).

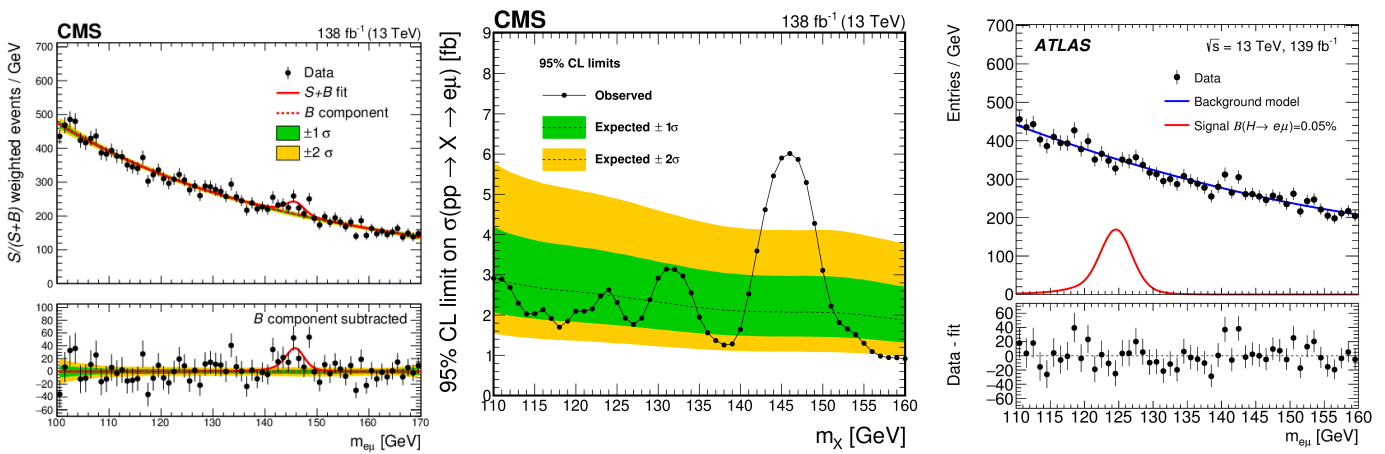

**Figure 20.** **Left**: Observed $e\mu$-invariant-mass distribution [27] (CC BY 4.0). **Middle**: Measured limits BR($H \rightarrow e\mu$) at 95% CL [27] (CC BY 4.0). **Right**: Observed $e\mu$-invariant-mass distribution [28] (CC BY 4.0).

### 2.6. Invisible Decays

A strong motivation for the search for invisible Higgs boson decays is that this search can give insight into the hidden sectors of the Higgs field [29]. Direct searches were made by the ATLAS Collaboration for Higgs bosons that decay into invisible particles (leading to missing energy) [30]. Signatures could be jet+$E_T^{miss}$, $VBF$+$E_T^{miss}$+photon, $t\bar{t}$+$E_T^{miss}$, $Z(\ell\ell)$+$E_T^{miss}$, and $VBF$+$E_T^{miss}$ final states. The photon could arise from ISR QED radiation and jets from QCD radiation. No indication of a signal has been found, and a limit is set on the BR($H \to inv$) < 0.107 (0.077 exp.) at 95% CL, assuming a SM production cross-section [30] (Figure 21, left); here "$inv$" denotes the above listed final states of invisible particles. Figure 21, right, shows details of the limit setting and an interpretation as limits on a weakly interacting massive particle (WIMP) [30]. Results by the CMS Collaboration were obtained in the $t\bar{t}H$ search BR($H \to inv$) < 0.47 (0.40) at 95% CL, and in the combined $ggH$, $VBF$, $VH$, and $t\bar{t}H$ analyses ($H \to inv$) < 0.15 (0.08) at 95% CL [31] (Figure 22, right). Details on the limit setting and an interpretation as limits on a WIMP are shown in Figure 22, right [31]. For light mass, dark matter (DM) candidates in the range of 0.1 GeV and $m_H/2$ (Higgs mass) model-dependent exclusion limits are found to complement direct-detection experiments.

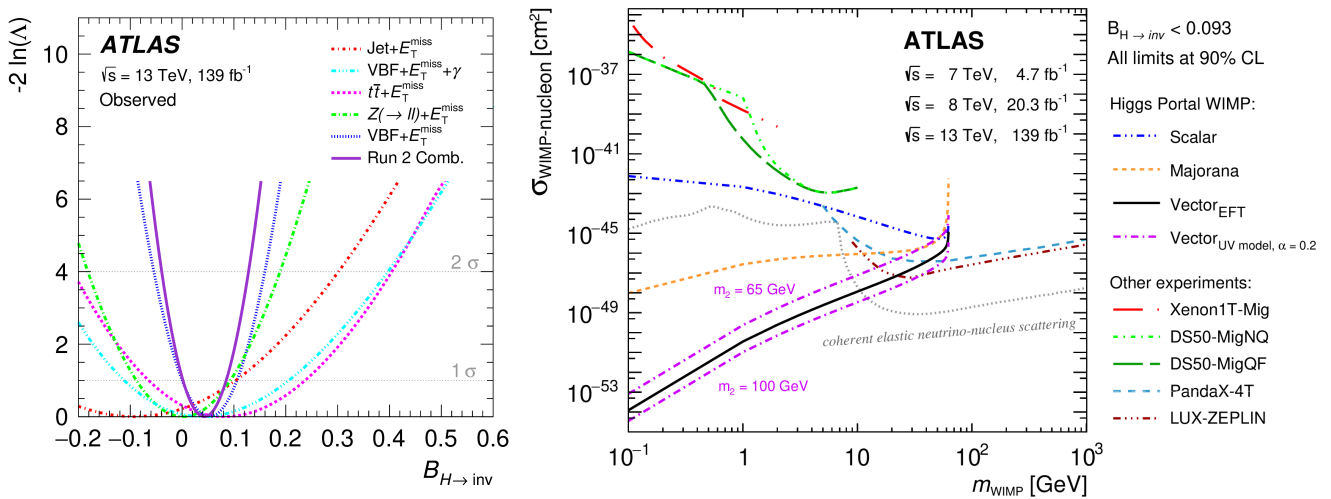

**Figure 21. Left**: Twice the negative logarithmic profile likelihood ratio as a function of BR($H \to inv$) decays into invisible particles [30] (CC BY 4.0). **Right**: Upper limits on the interaction cross-section of WIMP dark matter candidates with nucleons [30] (CC BY 4.0).

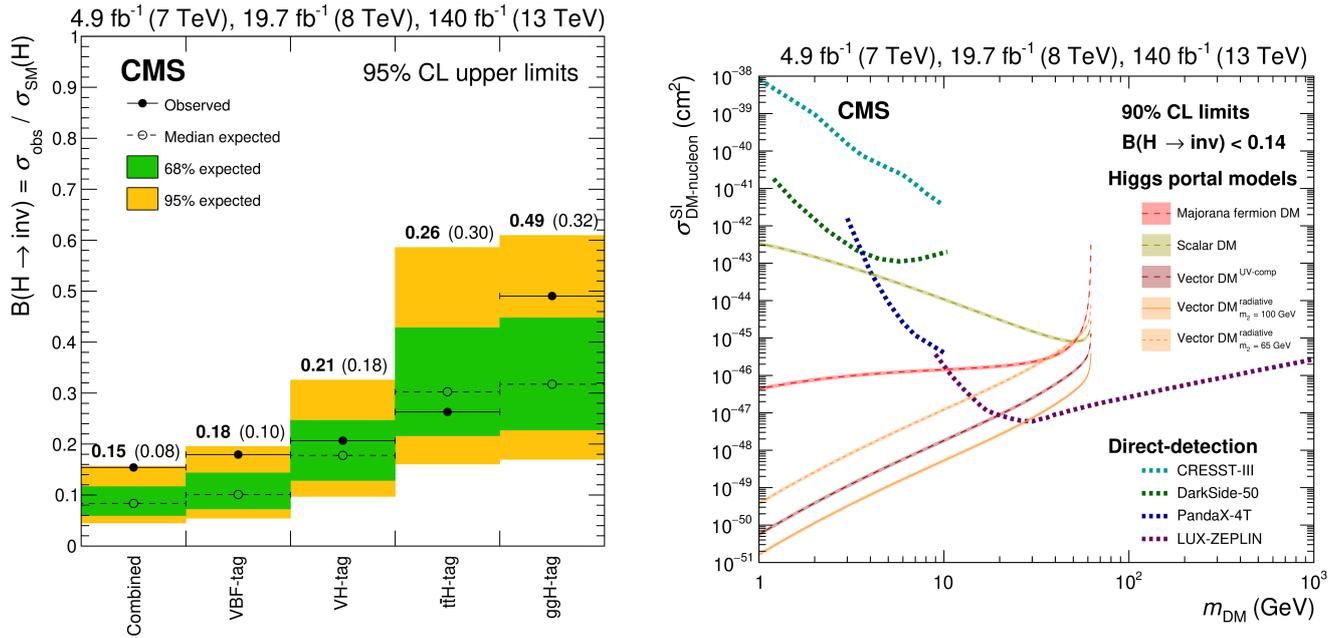

**Figure 22. Left**: Upper limits and Higgs boson decay branching ratios into invisible particles [31] (CC BY 4.0). **Right**: Upper limits on interaction cross-sections of dark matter (DM) candidates [31] (CC BY 4.0).

## 2.7. Dark Photons

There is strong astrophysical evidence suggesting the existence of DM [32]. The dark and visible physics sectors may interact through a portal, offering a potential experimental signature [33], including massive dark photons with energy $E_\gamma = m_{H/2}$ in the Higgs center-of-mass. Figure 23, left, shows the Feynman production diagram and Figure 23, right, shows the obtained upper limits on the Higgs boson decay branching ratios to $H \to \gamma\gamma_D$ [34], where the subscript "D" indicates the DM photon.

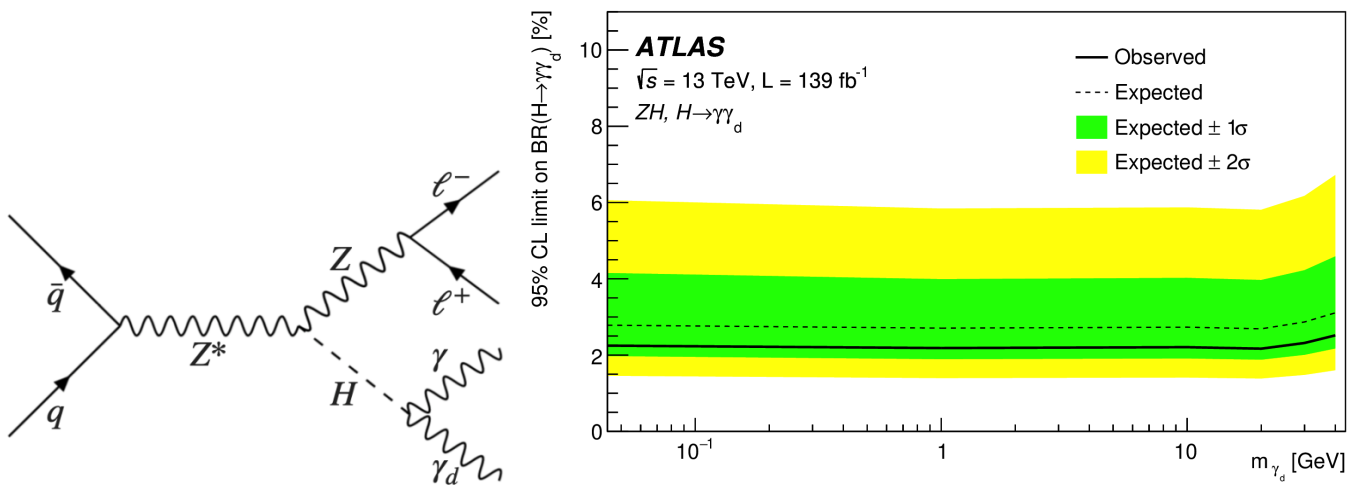

**Figure 23. Left**: Dark photon production Feynman diagram. **Right**: Upper limits on the Higgs boson decay branching ratios to $H \to \gamma\gamma_D$ [34] (CC BY 4.0).

## 2.8. Exotic Scalar Particles

An elegant solution to the strong *CP* problem by Roberto Peccei and Helen Quinn [35] is the introduction of a spontaneously broken U(1) symmetry leading to new light pseudoscalar particles –the axions– that couple very weakly to photons. At LHC, a search for the production of axion-like pseudoscalar particles in the decay $H \to Za$ with $Z \to \ell\ell$, where $\ell = e, \mu$ and $a \to \gamma\gamma$, is reported. Figure 24 shows simulated $ee\gamma\gamma$ and $\mu\mu\gamma\gamma$ signals (Figure 24, upper middle and upper right), efficiency times acceptance (Figure 24, lower left and lower middle), and limits on the Higgs boson decay branching fraction (Figure 24, lower right). Limits on models involving axion-like particles, formulated as an effective field theory, were also reported in Ref. [36].

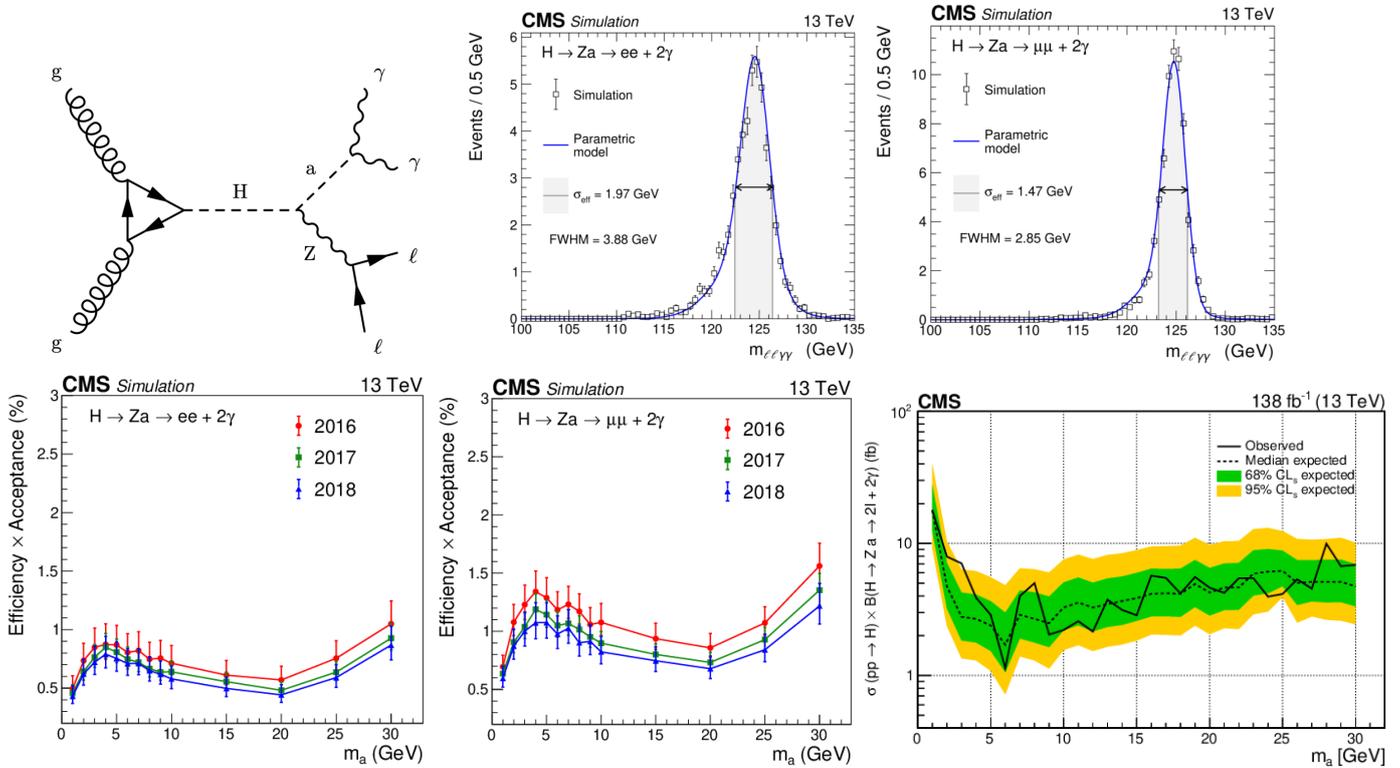

**Figure 24. Upper left**: Feynman diagram for $H \to Za$ with $Z \to \ell\ell$, where $\ell = e, \mu$ and $a \to \gamma\gamma$. **Upper middle** and **right**: Simulated $ee\gamma\gamma$ and $\mu\mu\gamma\gamma$ signals [36] (CC BY 4.0). **Lower**: Efficiency times acceptance (**left**) and (**middle**) and limits on the Higgs boson decay branching fraction (**right**) [36] (CC BY 4.0).

In models with two Higgs doublets extended with a scalar singlet (2HDM+S), the decay $H \to aa \to \mu\mu bb$ and $\tau\tau bb$ may exist. Figure 25, left, shows limits on the Higgs boson decay branching ratio $H \to aa \to \mu\mu bb$, $\tau\tau bb$ and Figure 25, right, the limits in the parameter space $(m_a, \tan\beta)$ [37].

In the 2HDM+S framework, low-mass di-muons are predicted $a \to \mu\mu$. For setting limits, $\tan\beta = 0.5$ is assumed, and $\sigma(pp \to a) = \sin^2(\Theta_H) \times B \times A$, where $B = BR(a \to \mu\mu)$, with acceptance $A$. No significant excess of events above the expectation from the SM background was observed [38], and the resulting limits are compared to previous limits from LHCb [39] and BaBar [40] experiments. Figure 26, left, compares the observed and the expected limits on the production cross-section times branching ratio times acceptance and Figure 26, right, shows the limit on the mixing angle $\sin\Theta_H$ [38].

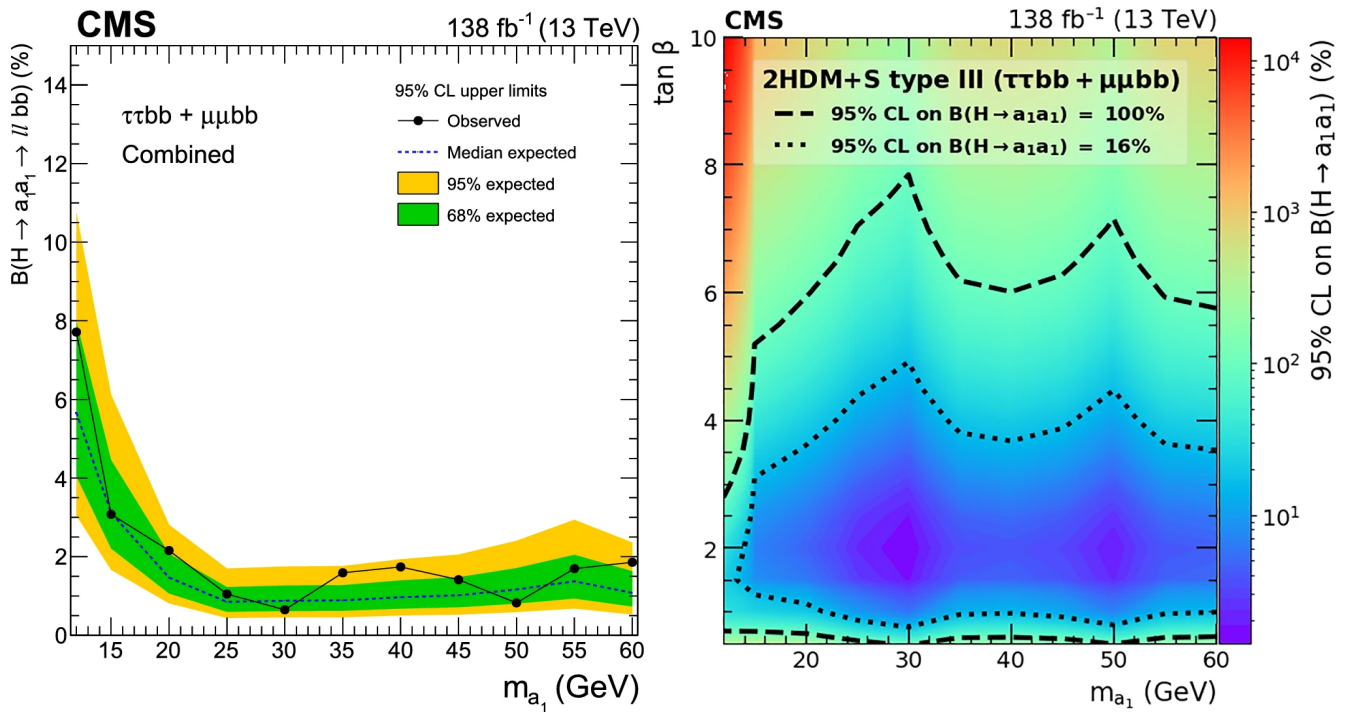

**Figure 25.** **Left**: Limits on the Higgs boson decay branching ratio $H \to aa \to \mu\mu bb, \tau\tau bb$ [37] (CC BY 4.0). **Right**: Limits in the parameter space $(m_a, \tan\beta)$ [37] (CC BY 4.0).

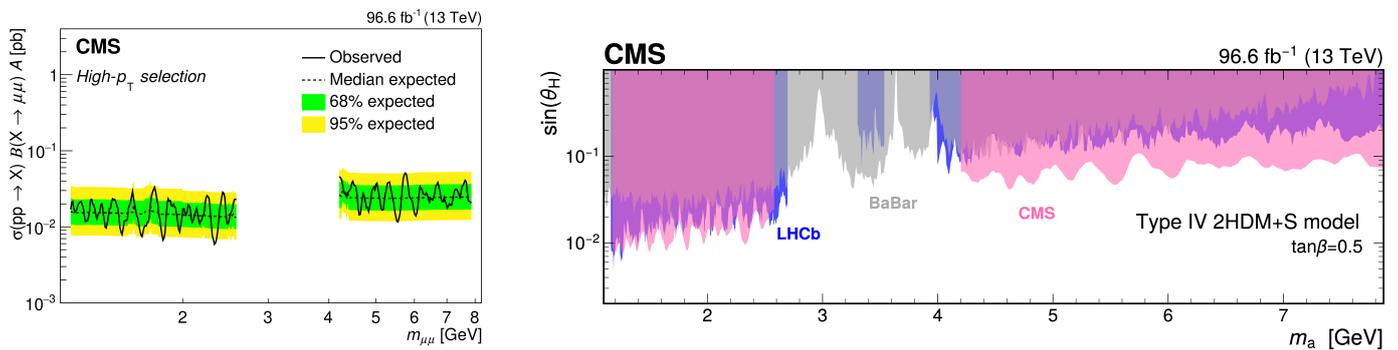

**Figure 26.** **Left**: Observed versus expected limits at 95% CL on the production cross-section times branching ratio times acceptance [38] (CC BY 4.0). **Right**: Limits at 90% CL on the mixing angle $\sin\Theta_H$ [38] (CC BY 4.0) compared to previous limits from LHCb [39] and BaBar [40] experiments.

Low-mass spin-0 particles can decay promptly to a closely separated photon pair that is reconstructed as a single photon-like object, $H \to AA \to 4\gamma$. Figure 27, left, compares the number of events for data, the simulated signal and background and Figure 27, right, represents the resulting limits on the branching ratio BR($H \to AA \to 4\gamma$) [41].

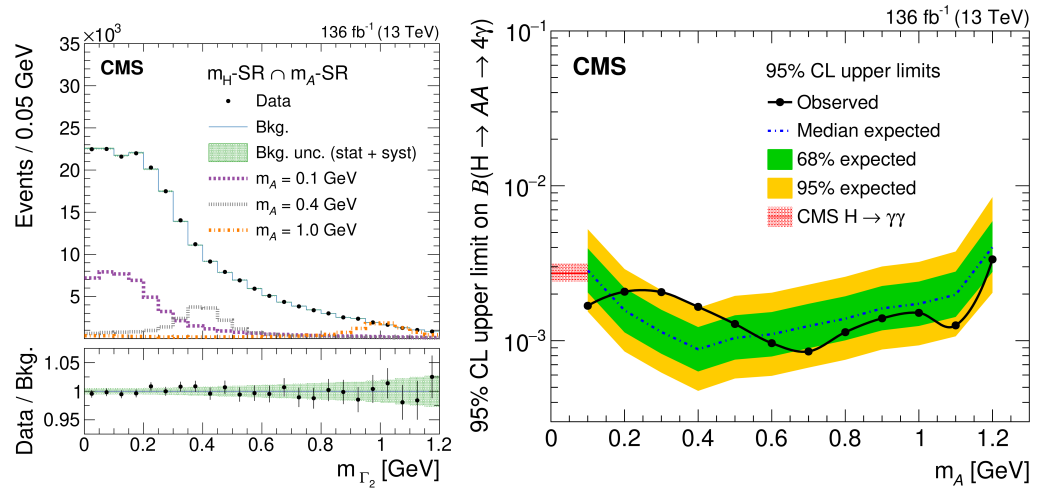

**Figure 27.** **Left**: Number of events for data, versus the simulated signal and background [41] (CC BY 4.0). **Right**: Limits on the Higgs boson decay branching ratio BR($H \to AA \to 4\gamma$) [41] (CC BY 4.0).

## 3. Enhanced Production Modes

BSM physics may enhance (or decrease) the production rates of Higgs bosons. This may lead to the observation of a SM reaction, where the SM predicts a rate below the sensitivity reach. Here, the measurements of $tH$ ($ttH$) and $HH$ signals are discussed.

### 3.1. Single Top Higgs

The expected production rate of $ttH$ and $tH$ events with $H \to bb$ is different for *CP*-even and *CP*-odd $ttH$ and $tH$ signals and, in particular, for the $tH$ enhancement in the production BSM for the *CP*-odd case. Figure 28, left, shows a $tH$ production Feynman diagram and Figure 28, right, the expected number of $ttH$ and $tH$ events for *CP*-even and *CP*-odd signals. The measured mixing angle between *CP*-even and *CP*-odd couplings is compatible with zero: $\alpha = 11^{+52}_{-72}$ degrees [42].

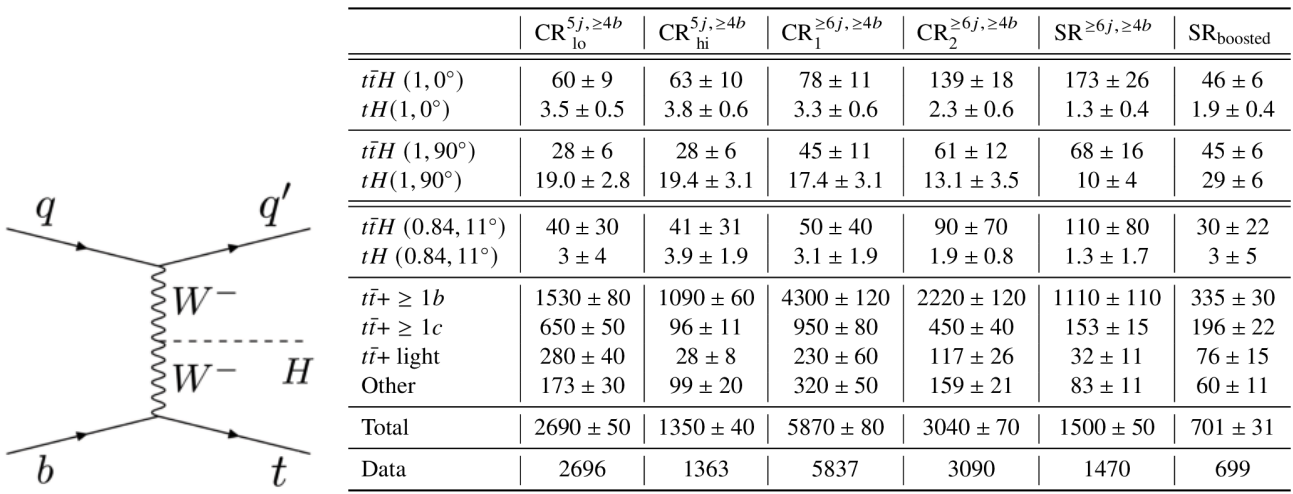

| | $\mathrm{CR}_{\mathrm{lo}}^{5j,\geq 4b}$ | $\mathrm{CR}_{\mathrm{hi}}^{5j,\geq 4b}$ | $\mathrm{CR}_1^{\geq 6j,\geq 4b}$ | $\mathrm{CR}_2^{\geq 6j,\geq 4b}$ | $\mathrm{SR}^{\geq 6j,\geq 4b}$ | $\mathrm{SR}_{\mathrm{boosted}}$ |
|---|---|---|---|---|---|---|
| $t\bar{t}H$ (1, 0°) | 60 ± 9 | 63 ± 10 | 78 ± 11 | 139 ± 18 | 173 ± 26 | 46 ± 6 |
| $tH$ (1, 0°) | 3.5 ± 0.5 | 3.8 ± 0.6 | 3.3 ± 0.6 | 2.3 ± 0.6 | 1.3 ± 0.4 | 1.9 ± 0.4 |
| $t\bar{t}H$ (1, 90°) | 28 ± 6 | 28 ± 6 | 45 ± 11 | 61 ± 12 | 45 ± 6 | |
| $tH$ (1, 90°) | 19.0 ± 2.8 | 19.4 ± 3.1 | 17.4 ± 3.1 | 13.1 ± 3.5 | 10 ± 4 | 29 ± 6 |
| $t\bar{t}H$ (0.84, 11°) | 40 ± 30 | 41 ± 31 | 50 ± 40 | 90 ± 70 | 110 ± 80 | 30 ± 22 |
| $tH$ (0.84, 11°) | 3 ± 4 | 3.9 ± 1.9 | 3.1 ± 1.9 | 1.9 ± 0.8 | 1.3 ± 1.7 | 3 ± 5 |
| $t\bar{t}+\geq 1b$ | 1530 ± 80 | 1090 ± 60 | 4300 ± 120 | 2220 ± 120 | 1110 ± 110 | 335 ± 30 |
| $t\bar{t}+\geq 1c$ | 650 ± 50 | 96 ± 11 | 950 ± 80 | 450 ± 40 | 153 ± 15 | 196 ± 22 |
| $t\bar{t}+$ light | 280 ± 40 | 28 ± 8 | 230 ± 60 | 117 ± 26 | 32 ± 11 | 76 ± 15 |
| Other | 173 ± 30 | 99 ± 20 | 320 ± 50 | 159 ± 21 | 83 ± 11 | 60 ± 11 |
| Total | 2690 ± 50 | 1350 ± 40 | 5870 ± 80 | 3040 ± 70 | 1500 ± 50 | 701 ± 31 |
| Data | 2696 | 1363 | 5837 | 3090 | 1470 | 699 |

**Figure 28.** **Left**: $tH$ production Feynman diagram. **Right**: ATLAS expected number of $ttH$ and $tH$ events for *CP*-even (0°) and *CP*-odd (90°) signals. The signals are given for unity coupling strength and post-fit coupling strength 0.84 [42].

The single top and Higgs boson production: $tH$, where $H \to WW/ZZ/\tau\tau$ and $H \to bb$, is combined with the production $ttH$ ($H \to \gamma\gamma$) [43]. Since the production is sensitive to the absolute values of the top quark Yukawa coupling, the Higgs boson cou-pling to vector bosons, $g(HVV)$, and, uniquely, their relative sign, are studied. The data favor $\kappa_t = 1.0$ over $\kappa_t = -1.0$ for the coupling modifier by more than 1.5 s.d. Figure 29, left, shows a distribution of data along with the simulated background and signal, for $\kappa_t = 1.0$ and $\kappa_V = 1.0$, in addition to the measured likelihood as a function of $\kappa_t$ (Figure 29, right).

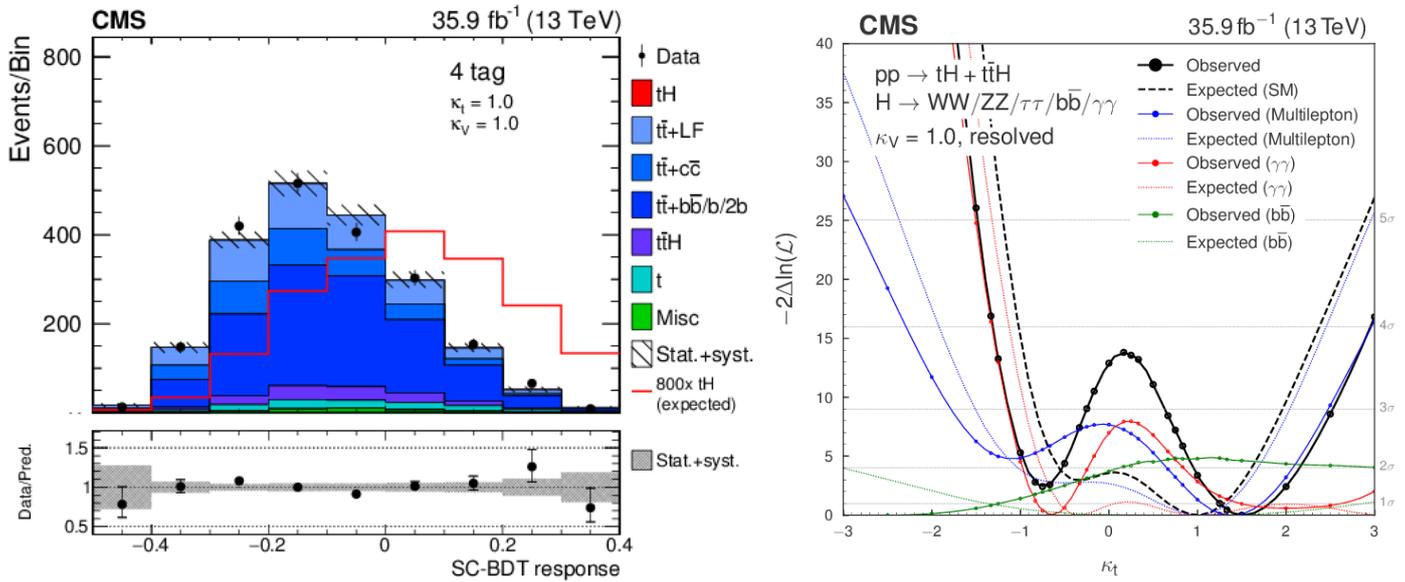

**Figure 29. Left**: Signal classification BDT (SC-BDT) for data, simulated background, and $tH$ signal for $\kappa_t = 1.0$ [43] (CC BY 4.0). **Right**: Measured likelihood as a function of $\kappa_t$ [43] (CC BY 4.0).

The single top and Higgs boson $tH$ production has also been analyzed with combined $H \to WW/ZZ/\tau\tau/bb/\gamma\gamma$ decays. The $t\bar{t}H$ normalization is kept fixed in the fit to the SM expectation, while the $tH$ signal strength is allowed to float. Discrepancies between observed and expected limits at $\kappa_t = 0.0$ are caused by the predicted $t\bar{t}H$ vanishing while the data favor even larger than expected yields for $t\bar{t}H$ production. Figure 30, left, compares the measured $tH$ and $t\bar{t}H$ couplings with those from the SM expectations and Figure 30, right, shows the measured likelihood as a function of $\kappa_t$ [44]. An upper limit at 95% CL on the $tH$ production rate of 14.6 times the SM expectation is observed, with an expectation of $19.3^{+9.2}_{-6.0}$ [44].

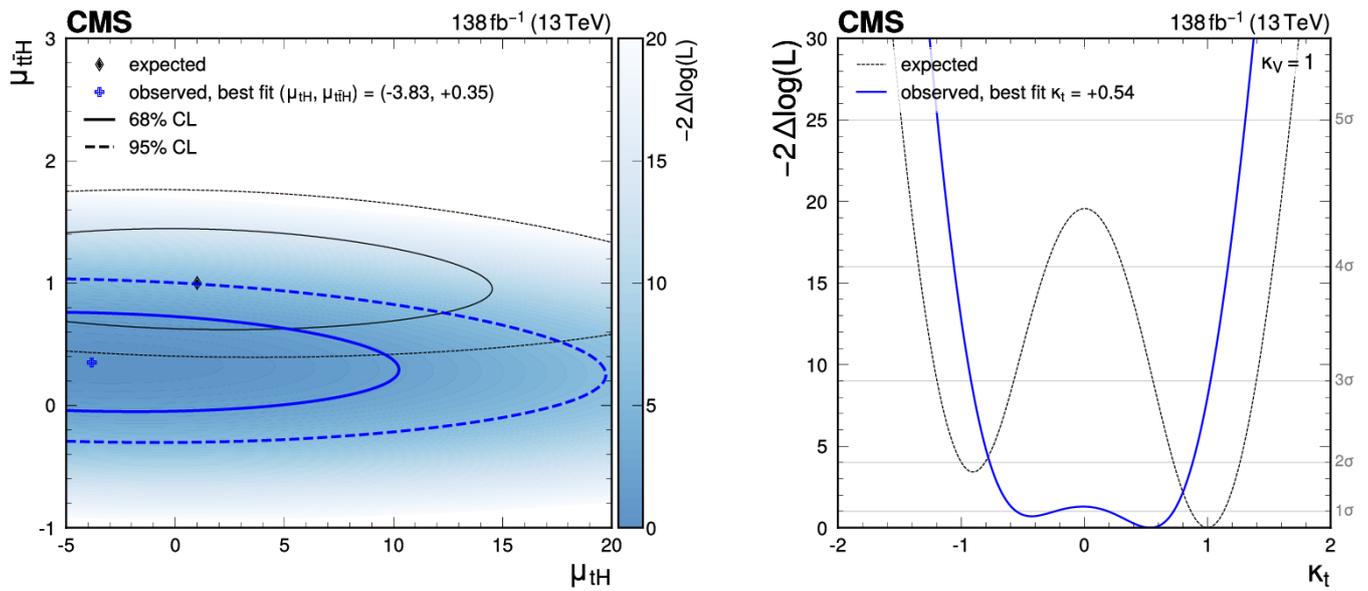

**Figure 30. Left**: Measured *tH* and *ttH* signal strengths compared to the SM expectations [44] (CC BY 4.0). **Right**: Measured likelihood as a function of $\kappa_t$. The SM value $\kappa_t = 1.0$ is outside the two standard deviation confidence interval [44] (CC BY 4.0).

### 3.2. Di-Higgs, $\kappa_\lambda$, $\kappa_{2V}$

The parameter $\kappa_\lambda$ is the *HHH* trilinear coupling modifier. Its value is left free in the fit to the recorded data, and in the SM, $\kappa_\lambda = 1$. Upper limits are set on the production signal strength, including the data from several final states $HH \to bbbb$, $bb\tau\tau$, $bb\gamma\gamma$ at 2.4 (2.9 exp.) and on the coupling modifier $-0.4 < \kappa_\lambda < 6.3$ at 95% CL [45,46]. Figure 31, upper, shows the dominant Feynman diagrams of *HH* production, while Figure 31, lower left, shows the upper limits on the *HH* signal strength with respect to the SM *ggF+VBF* induced production, and Figure 31, lower right, the upper limits on the cross-section of the *ggF* and *VBF* productions of *HH* [45].

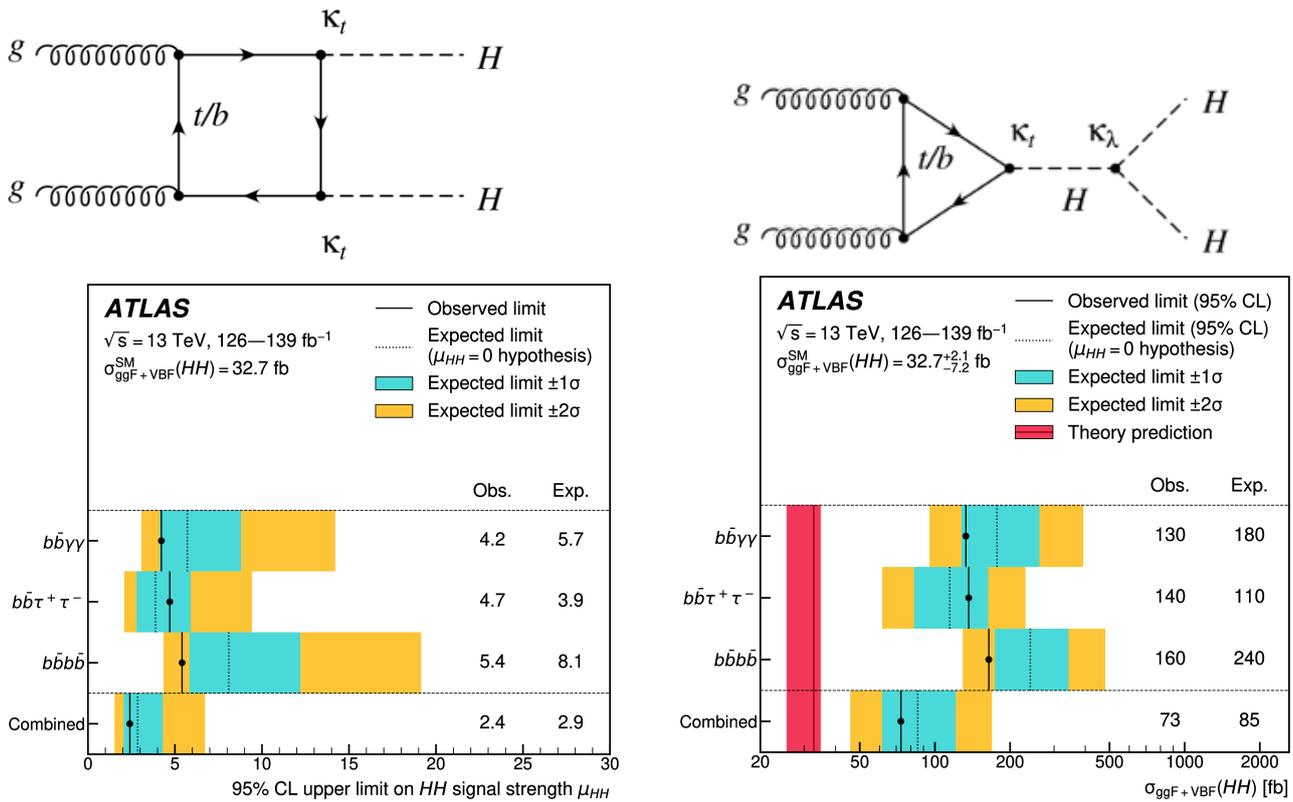

**Figure 31.** **Upper**: Feynman diagrams of $HH$ production. **Lower**: Upper limits (**left**) on the $HH$ signal strength with respect to the SM $ggF + VBF$ induced production and (**right**) on the cross-section of the $ggF$ and $VBF$ productions of $HH$ [45] (CC BY 4.0).

Observed and expected limits on the $HHH$ coupling modifier are set at 95% CL based on only the $bbbb$ final state and considering both $\kappa_\lambda$ in $[-3.5, 11.3]$ ($[-5.4, 11.4]$ exp.) and $\kappa_{2V}$ in $[-0.0, 2.1]$ ($[-0.1, 2.1]$ exp.) intervals. Limits are also set on the production cross-section in seven Higgs effective field theory benchmark scenarios [47]. Figure 32, upper left, shows a Feynman diagram for $VVHH$ scattering. Figure 32, upper middle and upper right, shows the data compared to simulated background and $HH$ signal and the limits on the $\kappa_\lambda$ and $\kappa_{2V}$, respectively. Figure 32, lower left and right, shows the limits on the $ggF + VBF$ $HH$ production cross-sections as a function of $\kappa_\lambda$ and $\kappa_{2V}$, respectively. The SM predictions $\kappa_\lambda = 1$ and $\kappa_{2V} = 1$ are in the measured ranges.

The production of $HH \rightarrow bbWW$ is constrained to be less than 14 (18 exp.) times the SM expectation at 95% CL [48]. In the Higgs effective field theory (HEFT) parameterization, limits are set for benchmark scenarios. In addition, the production $HH \rightarrow bbZZ$, *multilepton*, $bb\gamma\gamma$, $bb\tau\tau$, $bbbb$ is constrained at 3.4 (2.5 exp.) times the SM expectation at 95% CL [12]. Figure 33, left, shows limits on the $HH$ production cross-section for benchmark scenarios [48], in agreement with the expectations, along with the combined limits on the $HH$ production cross-section compared to the SM expectation in Figure 33, right [12].

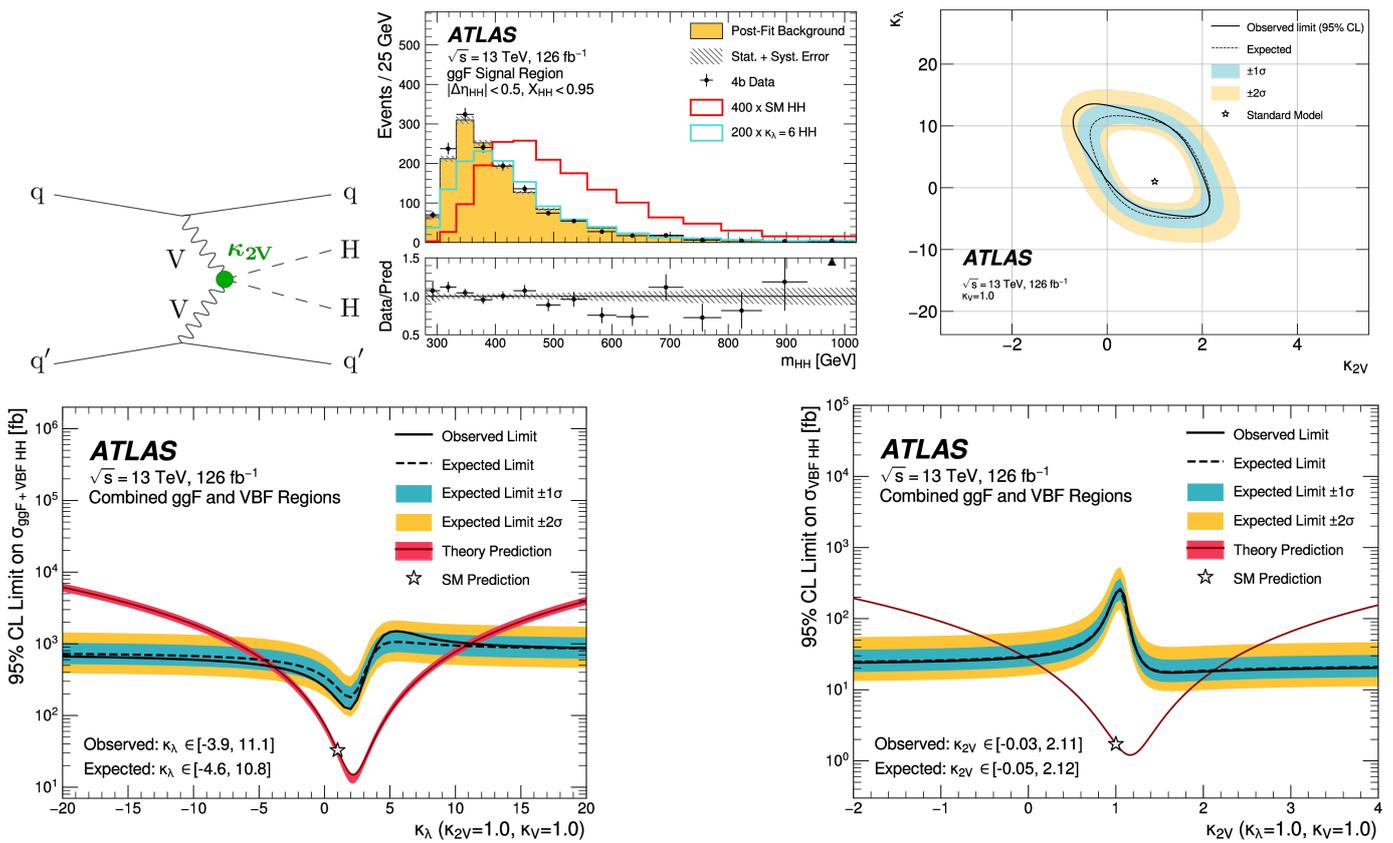

**Figure 32.** **Upper left**: Feynman diagram for *VVHH* scattering. **Upper middle**: Data, simulated background, and *HH* signal [47] (CC BY 4.0). **Upper right**: Limits in the $\kappa_\lambda$ and $\kappa_{2V}$ plane [47] (CC BY 4.0). **Lower**: Limits on the *VBF* and *ggF+VBF HH* production cross-sections as a function of (**left**) $\kappa_\lambda$ and (**right**) $\kappa_{2V}$ [47] (CC BY 4.0). See text for details.

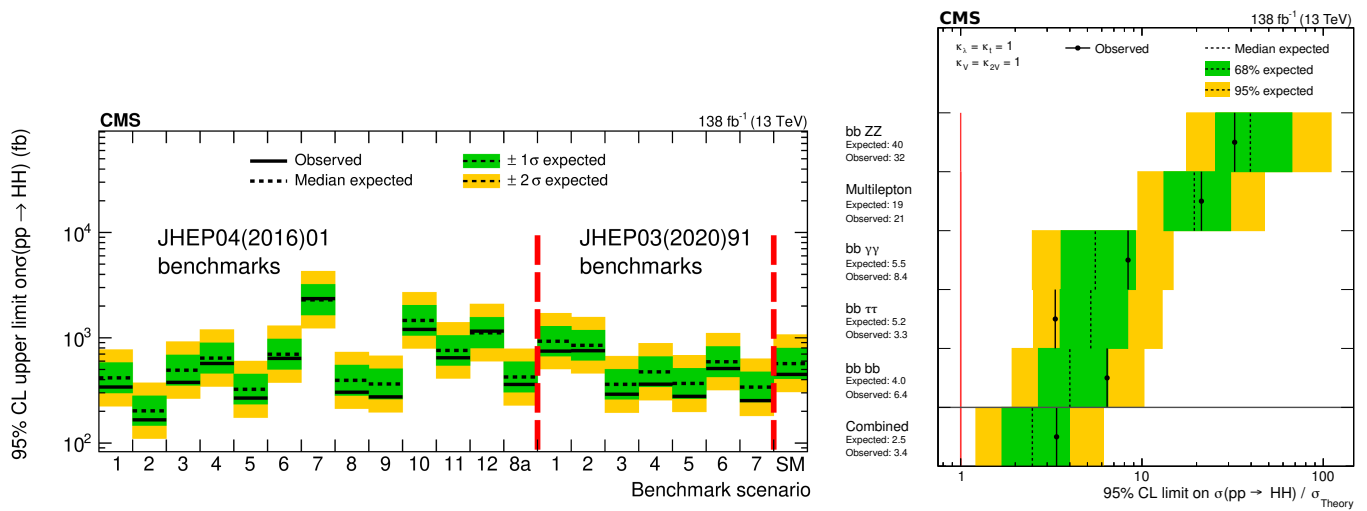

**Figure 33.** **Left**: Limits on the *HH* production cross-section for benchmark JEHP03, JHEP04 and SM scenarios [48] (CC BY 4.0). **Right**: Individual and combined limits on the *HH* production cross-section compared to the SM expectation [12] (CC BY 4.0).

An overview of the $\kappa_\lambda = 1.7^{+2.8}_{-1.7}$ and $\kappa_{2V} = 1.0^{+0.2}_{-0.2}$ measurements is given in Figure 34 [49], left and right, respectively. $\kappa_{2V} = 0$ is excluded with 6.6 s.d., and this establishes the existence of the quartic coupling VVHH, assuming $\kappa_\lambda = 1$. Values of $\kappa_\lambda$ outside the interval $-1.2 < \kappa_\lambda < 7.5$ are excluded at 95% CL [50].

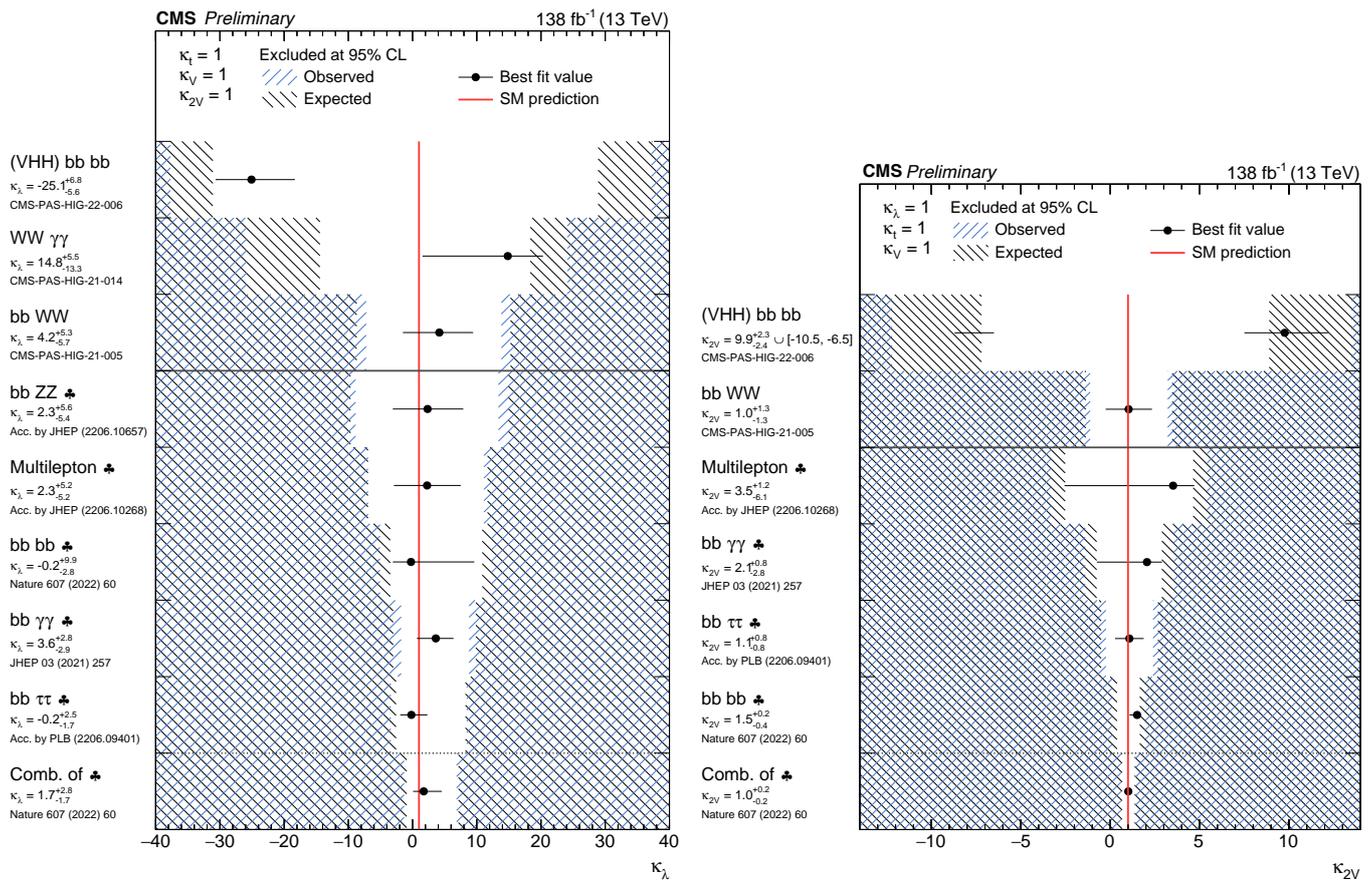

**Figure 34.** $\kappa_\lambda$ (**left**) and $\kappa_{2V}$ (**right**) measurements [49] (CC BY 4.0).

## 4. Higgs Boson Resonances

### 4.1. $X \to SH \to WW\tau\tau, ZZ\tau\tau$

A search for a new heavy scalar $X$ decaying into a Higgs boson and a new scalar singlet $S$, $X \to SH$, was performed in the mass range 500 GeV–1500 GeV for $X$ and in the mass range 200 GeV–500 GeV for $S$. The Higgs boson decays into a pair of $\tau$ leptons, and the $S$ decays into pairs $WW$ and $ZZ$, which subsequently decay into light leptons. No signal has been observed [51]. Figure 35, left, shows a Feynman diagram for the $SH$ production, while Figure 35, middle and right, shows simulated background plus $SH$ signal as a function of the BDT score for $WW$ and $ZZ$, respectively.

### 4.2. $X \to YH \to bbbb$

The search for the process $X \to YH \to bbbb$ has been performed for the mass ranges 0.9 TeV–4.0 TeV for $X$ and 60 GeV–600 GeV for $Y$ scalars [52]. For the interpretations, there are the two-real-scalar-singlet extension of the SM (TRSM) and the next-to-minimal model (NMSSM). Figure 36 [52] shows the invariant mass, $M_{JJ}$, the distribution of the two leading-$p_T$ jets in the event for the data, the simulated background and $X \to YH \to bbbb$ signals (Figure 36, left), and the observed exclusion limits in the mass parameter space $M_X$ and $M_Y$ at 95% CL (Figure 36, right). Both $Y$ and $H$ are reconstructed as Lorentz-boosted single large-area substructure jets J.

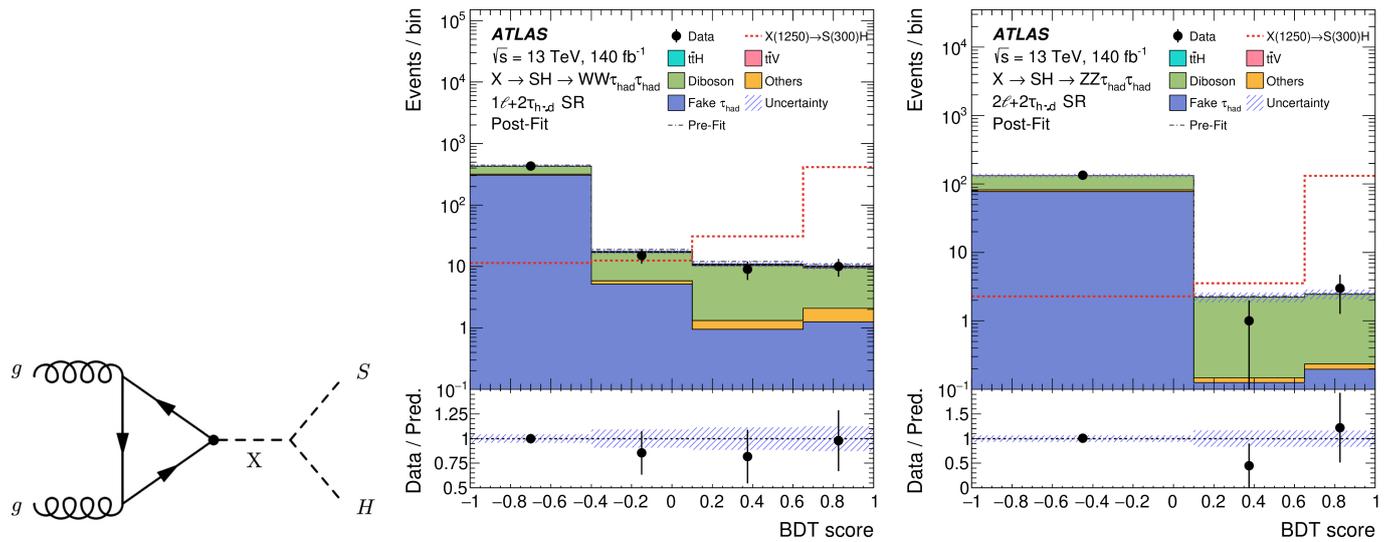

**Figure 35.** Feynman diagram for the $SH$ production (**left**); data, simulated background, and $SH$ signal as a function of the BDT score for $X \to SH \to WW\tau\tau$ (**middle**) and $X \to SH \to ZZ\tau\tau$ (**right**) [51] (CC BY 4.0).

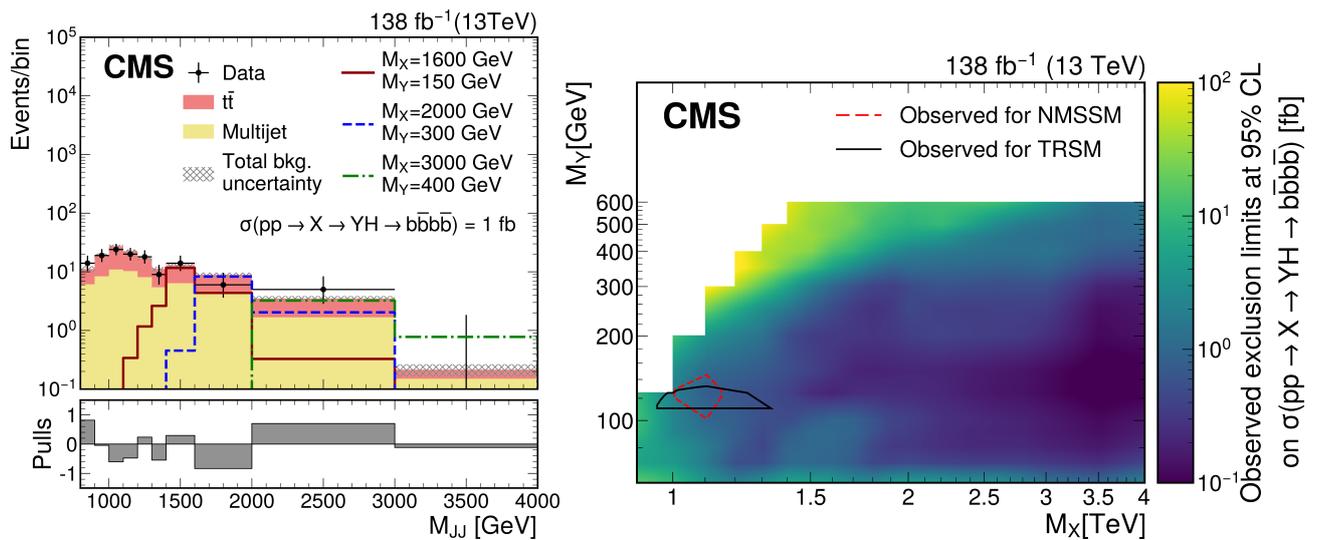

**Figure 36. Left**: Data, simulated background, and $X \to YH \to bbbb$ signals [52] (CC BY 4.0). Both $Y$ and $H$ are reconstructed as Lorentz-boosted single large-area substructure jets J. **Right**: Observed exclusion limits in the mass parameter space $M_X$ and $M_Y$ at 95% CL [52] (CC BY 4.0).

### 4.3. $X \to HH$

A summary of $X \to HH$ resonance searches and resulting limits are given in Figure 37 [53,54]. The ATLAS Collaboration sets upper limits on the production cross-section of a heavy scalar resonance decaying to two SM Higgs bosons at 95% CL between 0.96 fb and 600 fb (1.2 fb and 390 fb) in observation (expectation) [53], while the CMS Collaboration sets the limit between 0.1 fb and 150 fb, depending on the $X$ mass [54].

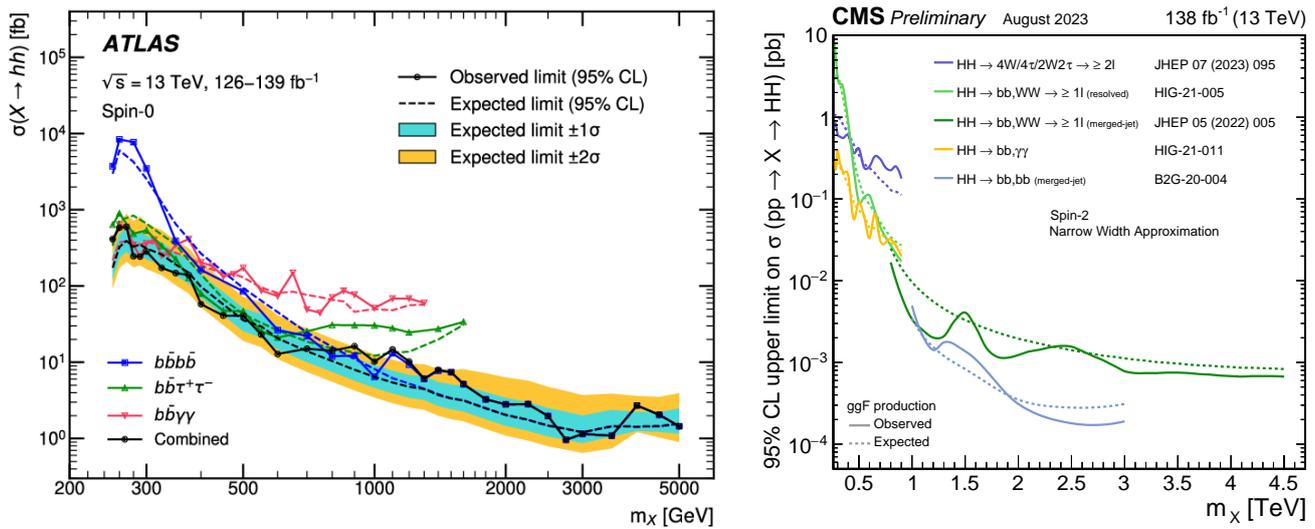

**Figure 37.** Summary of $X \to HH$ resonance searches and resulting limits by the (**left**) ATLAS [53] (CC BY 4.0) and (**right**) CMS [54] (CC BY 4.0) experiments.

## 5. Additional Neutral Higgs Bosons

### 5.1. SM-like Higgs Boson in the 70 GeV to 110 GeV Mass Range $H \to \gamma\gamma$ Decay

The search for a SM-like Higgs boson in the 70 GeV to 110 GeV mass range $H \to \gamma\gamma$ was performed in 36.3 fb$^{-1}$, 41.5 fb$^{-1}$ and 54.4 fb$^{-1}$ data from 2016, 2017, and 2018 data samples, respectively. The observed upper limit for the combined dataset ranged from 73 fb to 15 fb [55]. An excess with 2.9 s.d. local (1.3 s.d. global) significance was observed for a mass hypothesis of 95.4 GeV. Figure 38 shows the di-photon invariant mass distribution (Figure 38, left), resulting limits on the SM Higgs boson production times decay branching ratio into di-photons (Figure 38, middle), and local probability $p$-values as a function of the Higgs boson mass (Figure 38, right) [55]. No significant excess was observed by the ATLAS Collaboration [56] with 1.7 s.d. at 95.4 GeV. Earlier, an excess in combined LEP data was noted in the mass range of 90 GeV–100 GeV [57] of 2.3 s.d.; such an excess may be accommodated in a MSSM parameter scan [58].

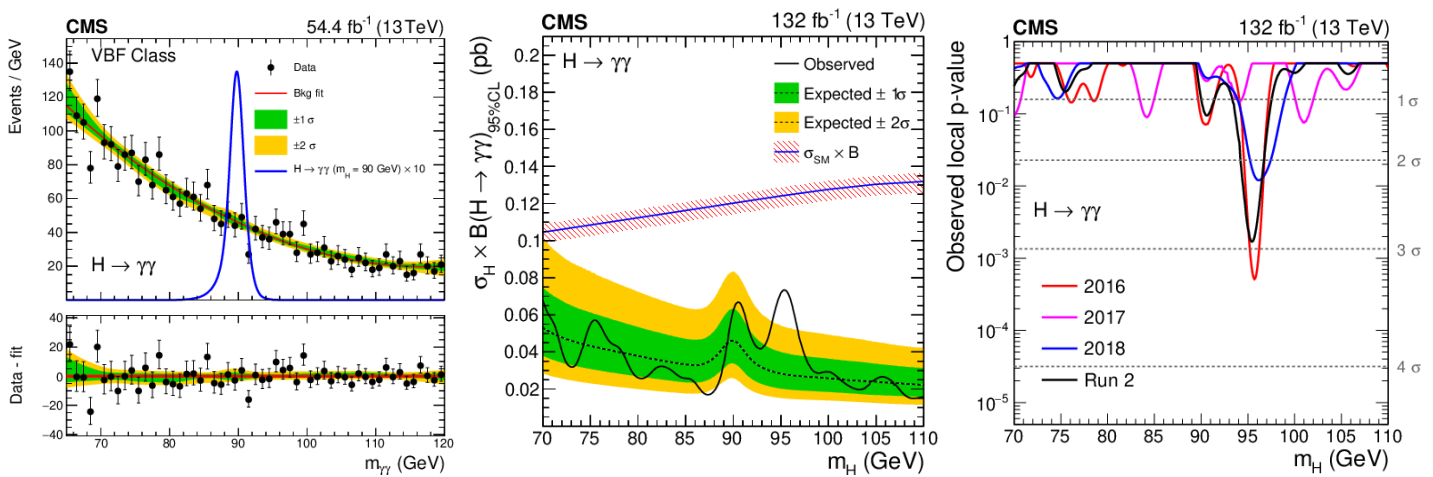

**Figure 38. Left**: Di-photon invariant mass distribution [55] (CC BY 4.0). **Middle**: Limits on the SM Higgs boson production times decay branching ratio into di-photons [55] (CC BY 4.0). **Right**: Local probability $p$-values as a function of the Higgs boson mass [55] (CC BY 4.0).

### 5.2. Heavy Higgs Bosons in $pp \to A \to HZ$

In the 2HDM, $m_A > m_H$ as well as the mass ranges $m_A > 800$ GeV, $m_H > 2m_t = 350$ GeV have been investigated [59]. The search final states are $t\bar{t}\ell\ell$ and $bb\nu\nu$. Figure 39, upper left and right, shows the data, as well as the simulated background and signal distributions, for $t\bar{t}\ell\ell$ and $bb\nu\nu$ final states, respectively, and Figure 39, lower, the resulting limits at 95% CL in the $(m_A, m_H)$ mass parameter space [59].

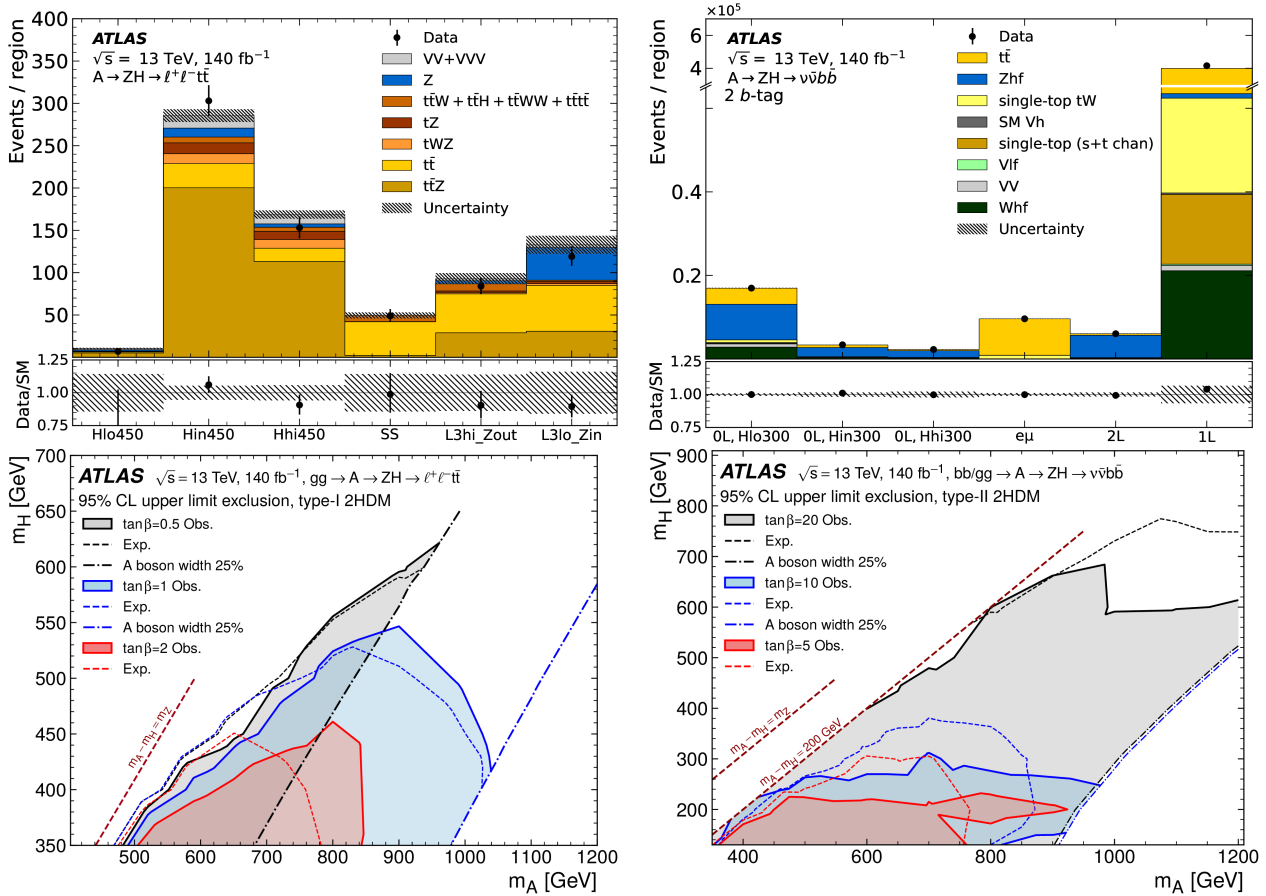

**Figure 39. Upper**: Data, simulated background and signal distributions for $t\bar{t}\ell\ell$ (**left**) and $bb\nu\nu$ (**right**) final states. **Lower**: Resulting limits at 95% CL in the $(m_A, m_H)$ parameter space [59] (CC BY 4.0).

### 5.3. Heavy Higgs Bosons in $pp \to VHH$

A search for $pp \to WHH$ and $pp \to ZHH$ was performed at LHC. This resulted in limits at 95% CL on $37.7 < \kappa_\lambda < 37.2$ ($-30.1 < \kappa_\lambda < 28.9$) and $-12.2 < \kappa_{VV} < 13.5$ ($-7.64 < \kappa_{VV} < 8.90$) [60]. Figure 40, left, shows a Feynman diagram for the $pp \to WHH$ and $pp \to ZHH$ production, and a simulated $WWH$ signal, and limits on the production cross-section $pp \to VHH$ are shown in Figure 40, middle and right, respectively.

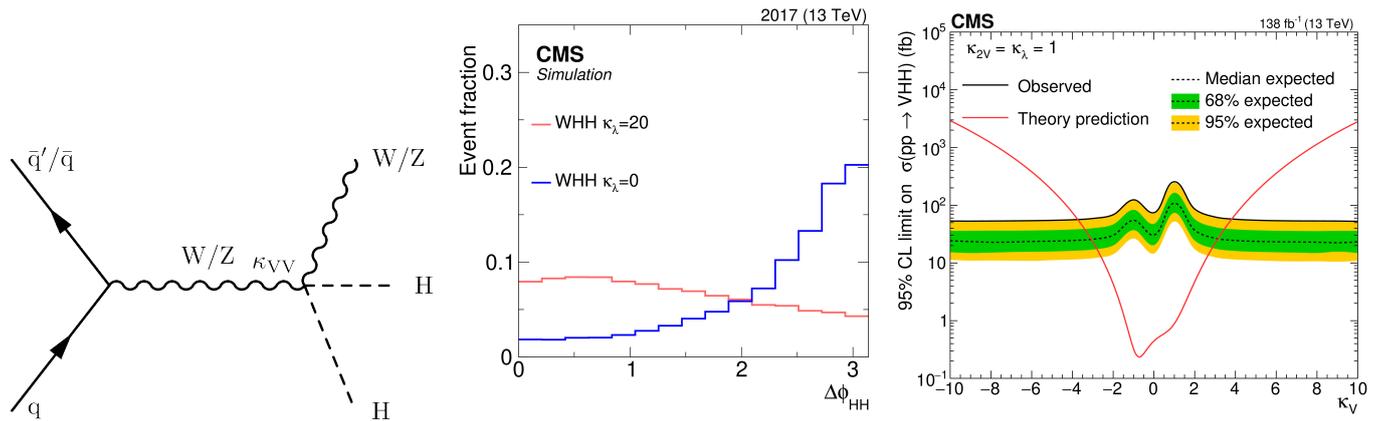

**Figure 40. Left**: Feynman diagram for the $pp \to WHH$ and $pp \to ZHH$ production. **Middle**: Simulated $WHH$ signal [60] (CC BY 4.0). **Right**: Limits on the production cross-section $pp \to VHH$ [60] (CC BY 4.0).

## 5.4. Heavy Higgs Bosons in $pp \to A \to VH, H \to hh$

A search for $VH, H \to hh$ was performed, where the vector boson decays $W \to \ell\nu$, $Z \to \ell\ell, \nu\nu$, with $\ell = e, \mu$. Interpretations have been given in the 2HDM with free parameters $m_A$, $m_H$, the ratio of the vacuum expectation values of the two Higgs doublets $\tan\beta$, and the Higgs boson coupling factor $\cos(\beta - \alpha)$ [61]. Figure 41 shows the Feynman diagram for the $VH, H \to hh$ production (Figure 41, upper left) along with the limits in the $(\cos(\beta - \alpha), m_A)$ parameter space and on the production cross-sections $WH$ and $ZH$ times branching ratio of $H \to hh \to bbbb$ (Figure 41, upper right, lower left and right, correspondingly) [61].

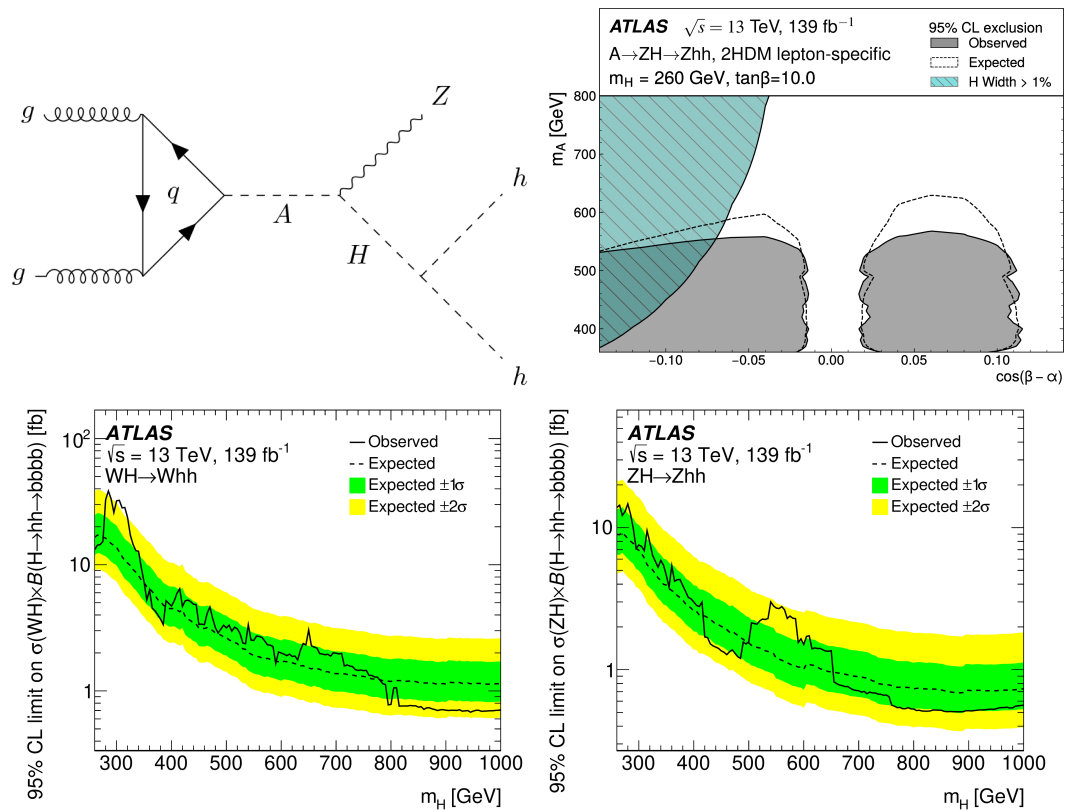

**Figure 41.** Feynman diagram for the $VH, H \to hh$ production (**upper left**) and the limits [61] (CC BY 4.0) in the $(\cos(\beta - \alpha), m_A)$ parameter space (see text for details) (**upper right**) and on the production cross-sections $WH$ (**lower left**) and $ZH$ (**lower right**) times branching ratio of $H \to hh \to bbbb$.

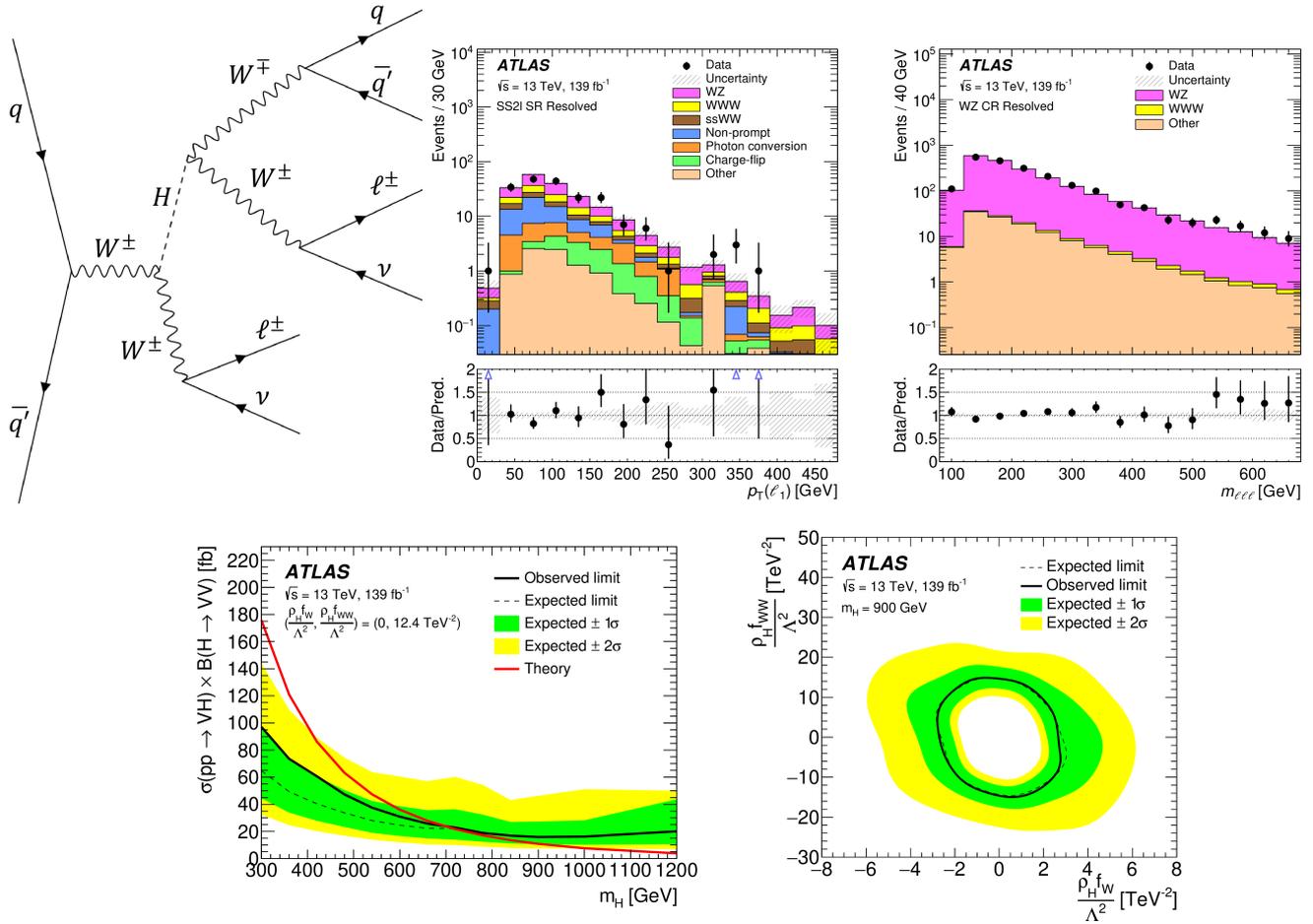

**Figure 42. Upper**: Feynman diagram for heavy Higgs boson production, with $H \to VV$ in same-sign two-lepton final states (**left**); Data, simulated background, and $H \to VV$ signals [63] (CC BY 4.0) for lepton transverse momentum (**middle**) and for three-lepton invariant mass distribution (**right**). **Lower**: Limits on the $pp \to VH \to H \to VV$ production cross-section (**left**) and in the parameter plane $(f_W, f_{WW})$ (**right**) [63] (CC BY 4.0). See text for details.

### 5.5. Heavy Higgs Bosons in $ttH/A \to tttt$: Multi-Lepton Final States

In the search for $ttH/A \to tttt$ multi-lepton final states, exactly two leptons with same-sign electric charges or at least three leptons are required [64]. In the 2HDM (Type-2), $\tan \beta < 1.2$ (0.5) are excluded for Higgs boson mass 400 (1000) GeV. Figure 43, upper left, shows a Feynman diagram for the $ttH/A \to tttt$ production, while the data, simulated background, and a $ttH/A \to tttt$ signal as a function of the BDT score, along with the limits on the $ttH/A$ production cross-section times the $H/A \to tt$ branching ratio and in the $(m_A, \tan \beta)$ are shown in Figure 43, upper right, lower left and lower right, respectively.

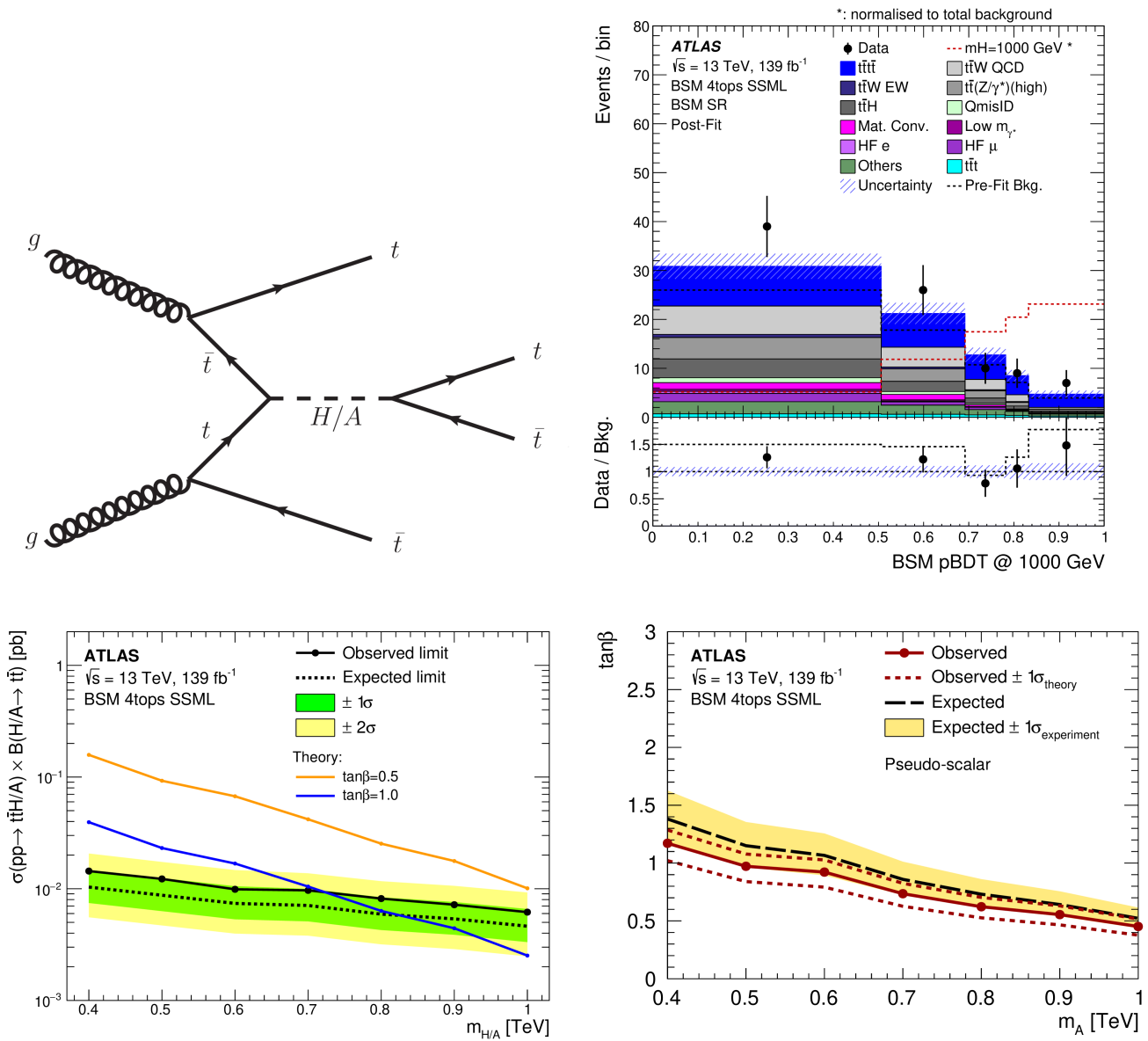

**Figure 43.** Feynman diagram for the $ttH/A \to tttt$ production (**upper left**); Data, simulated background and a $ttH/A \to tttt$ signal (**upper right**) along with limits on the $ttH/A$ production cross-section times the $H/A \to tt$ branching ratio (**lower left**) and exclusion limits in the $(m_A, \tan\beta)$ parameter planes (**lower right**) [64] (CC BY 4.0).

## 5.6. Heavy Higgs Bosons in $gg\Phi$, $bb\Phi$, $\Phi \to \tau\tau$

The search for a heavy neutral Higgs boson $\Phi$ with $\Phi \to \tau\tau$ was performed in $e\mu$, $e\tau_h$, $\mu\tau_h$, $\tau_h\tau_h$ final states [65]. Upper limits at 95% CL are set on the products of the branching fraction for the decay into tau leptons and the cross-sections for the production of a new boson $\Phi$, in addition to the H(125) boson, ranging from $\mathcal{O}(10\,\text{pb})$ for a mass of 60 GeV to 0.3 fb for a mass of 3.5 TeV. Figure 44 shows Feynman diagrams for the $gg\Phi$ (Figure 44, upper left) and $bb\Phi$ (Figure 44, upper right) production, the limits on the cross-section of $gg\Phi$ and $bb\Phi$ production times branching ratio for $\Phi \to \tau\tau$ (Figure 44, lower left and right, respectively) [65], along with the limits on benchmark scenarios of the MSSM that rely on the signal from three neutral Higgs bosons, one of which is associated with H(125). Expected and observed 95% CL exclusion contours are also set in the MSSM M125 and M125(EFT) scenarios, as shown in Figure 45 [65]. The data reveal two excesses, with local $p$-values equivalent to about three standard deviations at 0.1 and 1.2 TeV mass.

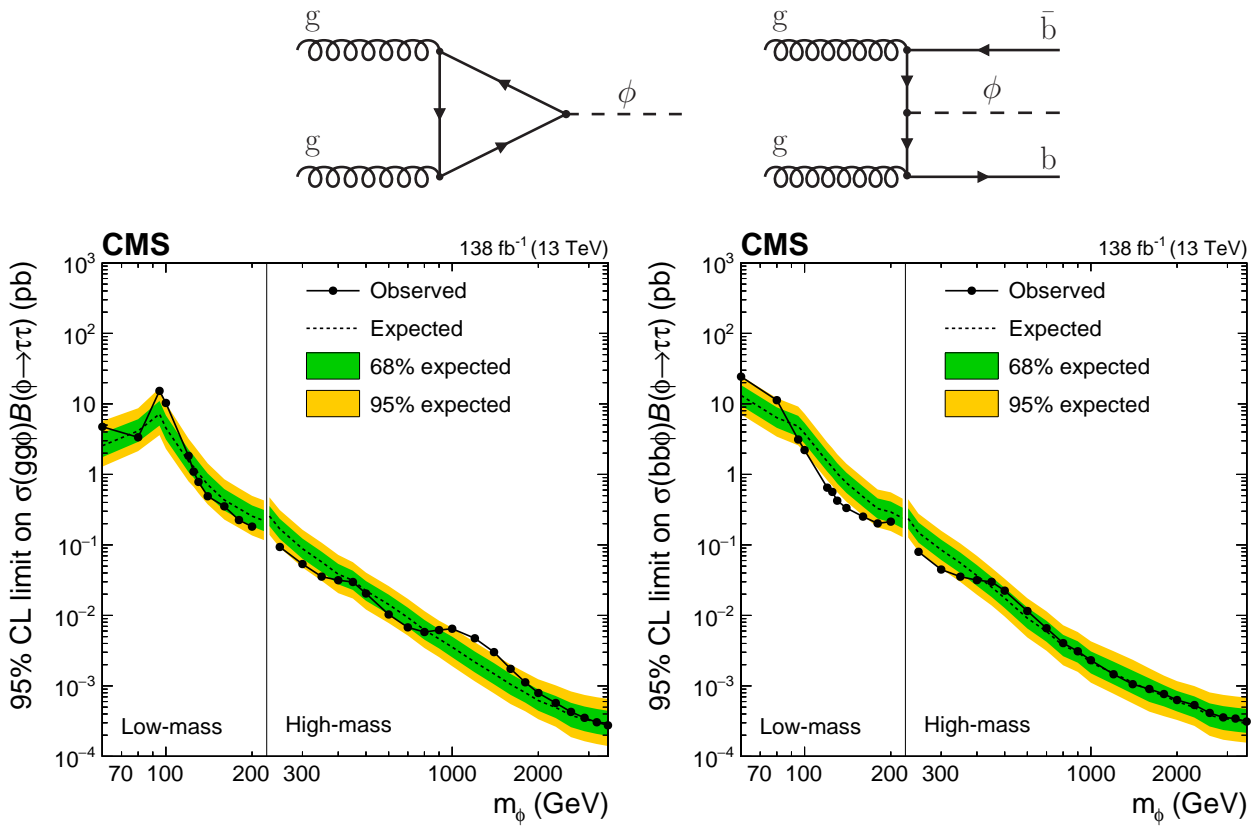

**Figure 44. Upper**: Feynman diagrams for the $gg\Phi$ (**left**) and $bb\Phi$ (**right**) production. **Lower**: limits on the cross-section for $gg\Phi$ (**left**) and $bb\Phi$ (**right**) production times branching ratios of $\Phi \to \tau\tau$ production [65] (CC BY 4.0).

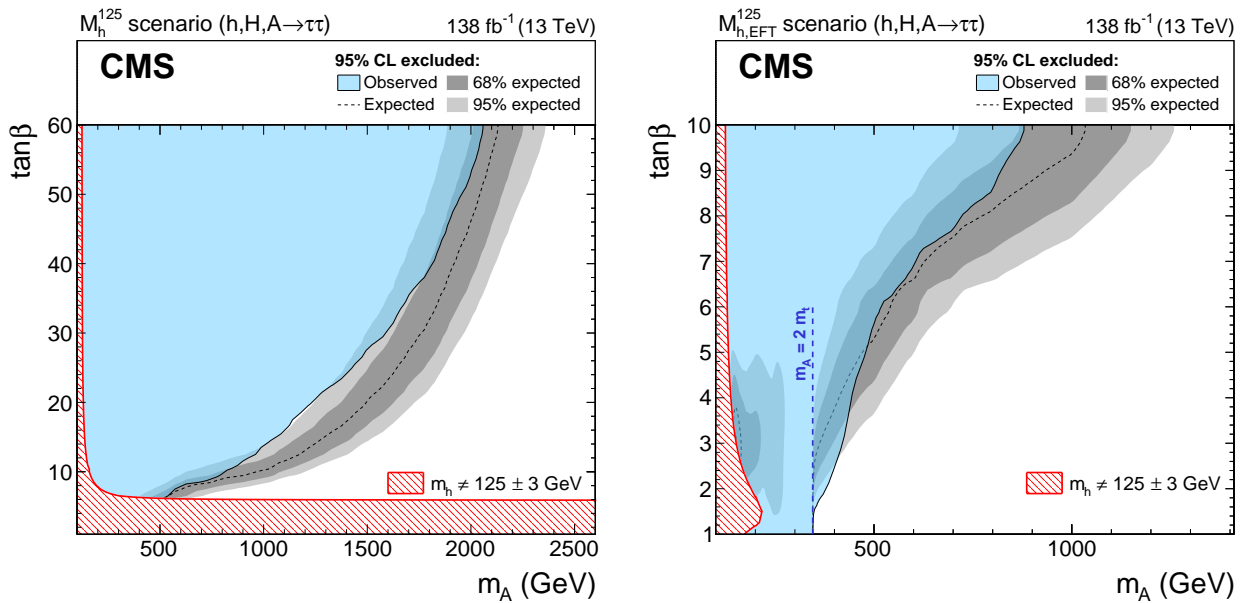

**Figure 45.** Expected and observed 95% CL exclusion contours for (**left**) MSSM M125 and (**right**) effective field theory M125(EFT) [65] (CC BY 4.0).

## 6. Higgs Bosons and Dark Matter

### 6.1. Dark Matter Produced in Association with a Higgs Boson Decaying into Taus

The search for DM considered to be produced in association with a Higgs boson decaying into $\tau$s was performed with the final state of two hadronically decaying tau leptons and MET, $E_T^{miss}$. For the interpretation, the 2HDM+$a$ was used [66]. Figure 46, upper, shows three production Feynman diagrams leading to DM and a pair of $\tau$s and Figure 46, lower left, middle and right, shows the data, simulated background, and a DM plus $\tau$s signal, along with the exclusion limits in the $(m_a, m_A)$ and $(m_A, \tan\beta)$ parameter planes, respectively.

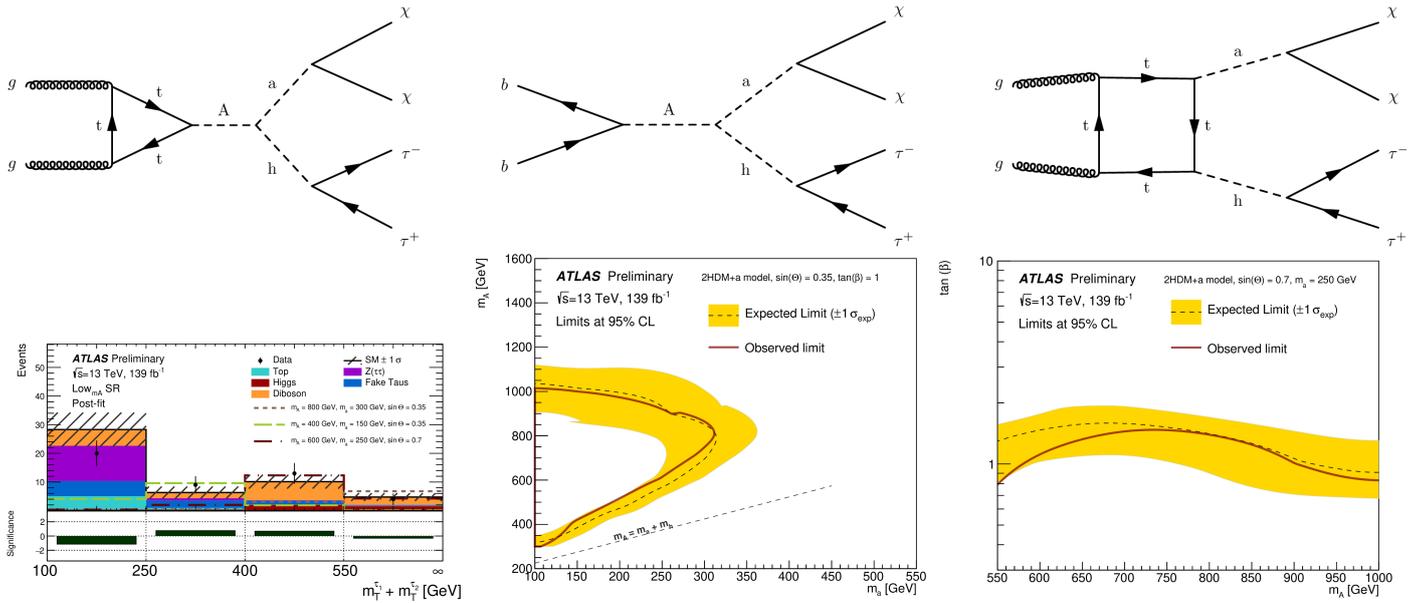

**Figure 46. Upper**: Production Feynman diagrams leading to DM and a pair of $\tau$s. **Lower**: Data, simulated background and a DM plus $\tau$s signal (**left**), exclusion limits in the $(m_a, m_A)$ parameter plane (**middle**), and in the $(m_A, \tan\beta)$ parameter planes (**right**) [66] (CC BY 4.0).

### 6.2. Dark Higgs Boson

In the Dark Abelian Higgs model, an additional U(1)$_D$ gauge symmetry can lead to a dark photon, $A'$, that mediates the dark sector interactions with the SM. The U(1)$_D$ symmetry group is spontaneously broken by a Higgs mechanism, leading to $A'$, which acquires a mass by adding a dark Higgs boson, $h_D$. The search for the process $Z \to A'h_D \to A'A'A^{(*)'}$ was performed, where two on-shell $A' \to \ell\ell$ with ($\ell = e$ or $\mu$) were required. The new results [67] were compared with results from the Belle experiment [68]. Figure 47 shows a Feynman diagram of the $Z \to A'h_D \to A'A'A^{(*)'}$ signal (Figure 47, left), data and simulated background, and $Z \to A'h_D \to A'A'A^{(*)'}$, where two on-shell $A' \to \ell\ell$, with a ($\ell = e$ or $\mu$) signal (Figure 47, middle) and the limits on the coupling as a function of $m_{A'}$ [67] (Figure 47, right).

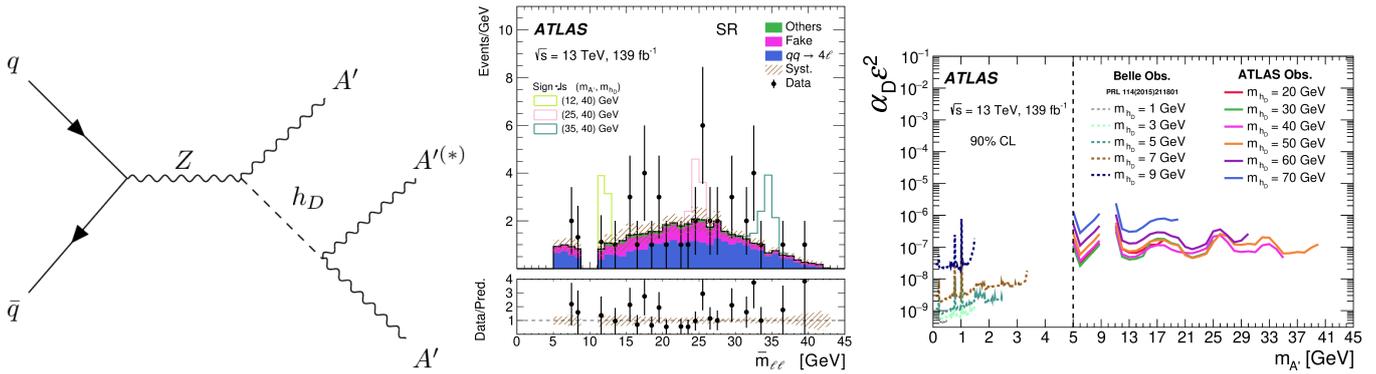

**Figure 47.** **Left**: Feynman diagram of the $Z \to A'h_D \to A'A'A^{(*)'}$ production. **Middle**: Data and simulated background as well as the $Z \to A'h_D \to A'A'A^{(*)'}$ signal, where two on-shell $A' \to \ell\ell$, with a ($\ell = e/\mu e$ or $\mu$) signal [67] (CC BY 4.0). **Right**: Limits on the coupling as a function of $m_{A'}$ [67] (CC BY 4.0).

### 6.3. Combination of Searches in the 2HDM and Dark Matter Searches

A combination of 2HDM and DM searches was performed at LHC. The most sensitive searches were $E_T^{miss}$ plus leptonically decaying $Z$ bosons, $E_T^{miss}$ plus $H \to bb$, and charged Higgs bosons with top and b-quarks decays [69]. Figure 48 shows the combined results as exclusion contours in the $(m_a, m_A)$ parameter space.

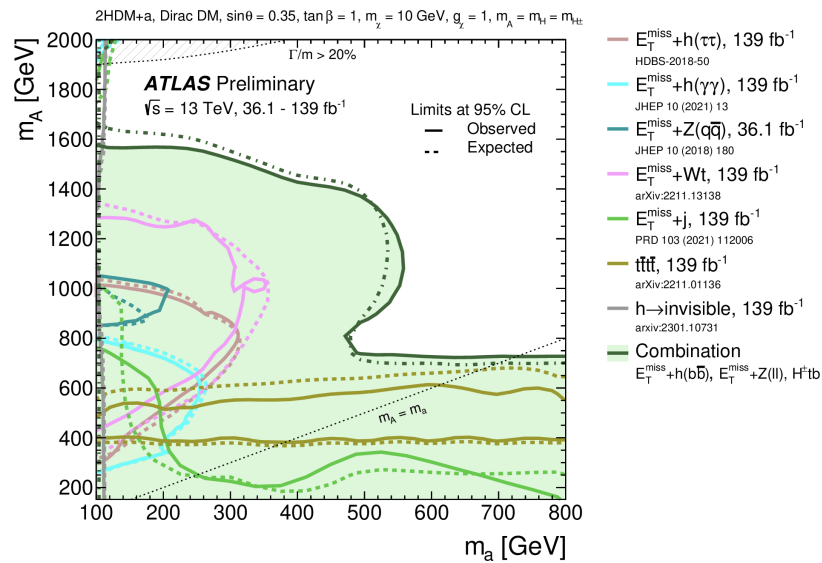

**Figure 48.** Results from combination of searches in the 2HDM and dark matter searches [69] (CC BY 4.0).

## 7. Additional Charged Higgs Bosons

### 7.1. $pp \to tt$ with $t \to H^\pm b$, $H^\pm \to W^\pm a$, where $a \to \mu\mu$

The search for the process $pp \to tt$ with $t \to H^\pm b$, $H^\pm \to W^\pm a$, where $a \to \mu\mu$ has been performed in the mass region $15 < m_a < 72$ GeV and $120 < m_{H^\pm} < 160$ GeV [70]. Figure 49 shows a Feynman diagram for the charged Higgs boson production (Figure 49, left) along with the data, simulated background, and signal for the di-muon invariant mass (Figure 49, middle), and limits on the top into charged Higgs boson b-quark decay branching ratio at 95% CL (Figure 49, right). No significant excess is observed, and the smallest local $p$-value is 0.008 at $m_a = 27$ GeV, corresponding to a local significance of about 2.4 s.d. The slight excess was observed at $m_a = 27$ GeV in both the $e\mu\mu$ and $\mu\mu\mu$ channels [70].

                                                                                  

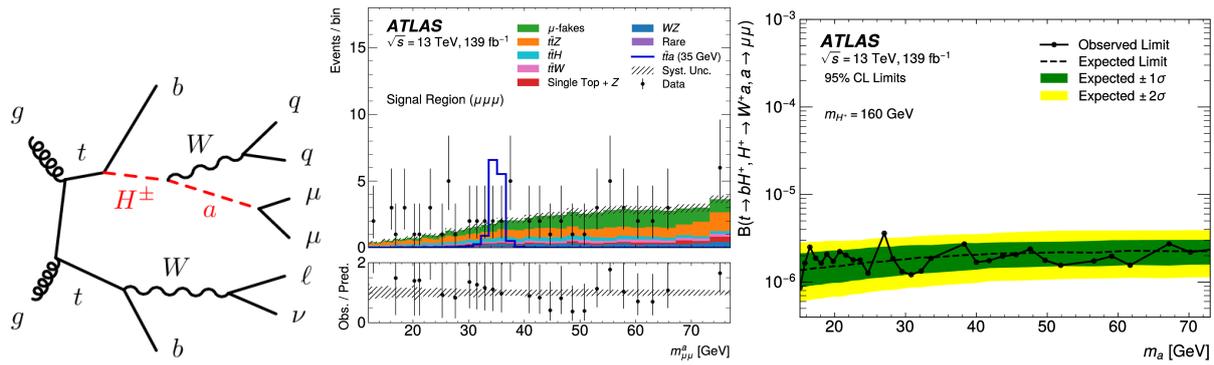

**Figure 49. Left**: Feynman diagram for the charged Higgs boson production. **Middle**: Data, simulated background and a signal for the di-muon invariant mass [70] (CC BY 4.0). **Right**: Limits on the top into charged Higgs boson and *b*-quark decay branching ratio at 95% CL [70] (CC BY 4.0).

### 7.2. $t \to H^\pm b, H^\pm \to cb$

The search for $t \to H^\pm b, H^\pm \to cb$ is based on an isolated electron or muon and at least four jets and three *b*-tags in 3HDM [71,72]. The top quark can decay into a lighter charged Higgs boson. The couplings *X*, *Y*, and *Z* are the Yukawa interactions of the lighter charged Higgs bosons. Those are functions of the Higgs-doublet vacuum expectation values and the mixing angle between the charged Higgs bosons [73]. Figure 50 shows a Feynman diagram of the $t \to H^\pm b, H^\pm \to cb$ production (Figure 50, left), data, simulated background, and a $t \to H^\pm b, H^\pm \to cb$ signal (Figure 50, middle), and the limits on the branching ratio $H^\pm \to cb$ as a function of the charged Higgs boson mass (Figure 50, right). The observed exclusion limits are consistently weaker than the expectation [73]. The largest excess in data has a local significance of about 3 s.d. for $m_{H^+} = 130$ GeV.

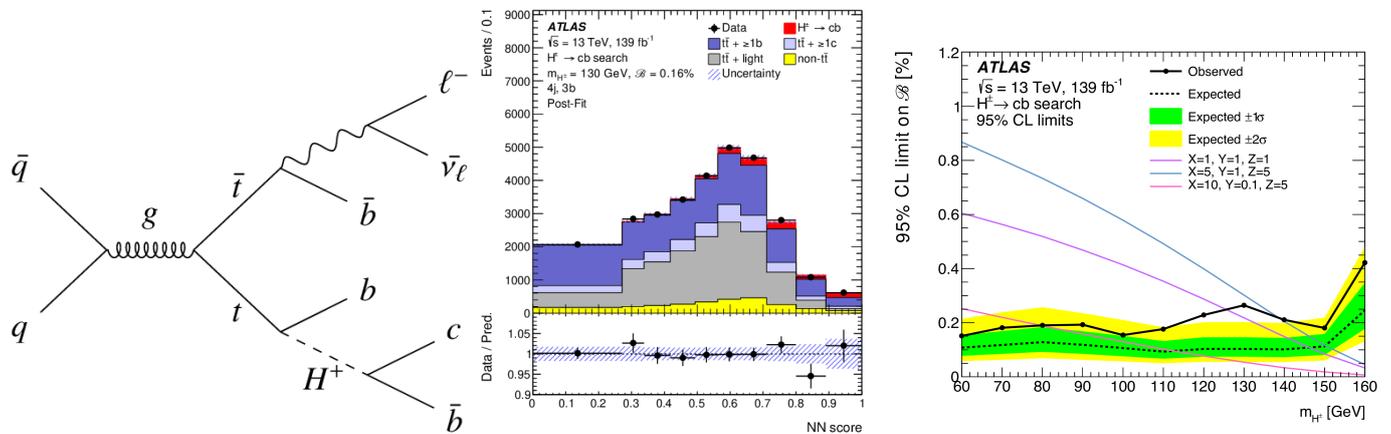

**Figure 50. Left**: Feynman diagram of the $t \to H^\pm b, H^\pm \to cb$ production. **Middle**: Data, simulated background, and a $t \to H^\pm b, H^\pm \to cb$ signal [73] (CC BY 4.0). **Right**: Limits on the branching ratio $H^\pm \to cb$ as a function of the charged Higgs boson mass [73] (CC BY 4.0).

### 7.3. $tbH^\pm, H^\pm \to WH, H \to \tau\tau$

A search for a charged Higgs boson $tbH^\pm, H^\pm \to WH, H \to \tau\tau$ decaying into a heavy neutral Higgs boson and a *W* boson was performed [74]. Signal processes were generated with four mass hypotheses $m_{H^+} = 300, 400, 500,$ and $700$ GeV, under the assumption that $m_h = 125$ GeV and $m_H = 200$ GeV. Upper limits on the $tbH^\pm, H^\pm \to WH, H \to \tau\tau$ production cross-section times branching ratio were set in the range 0.085 pb–0.019 pb for the charged Higgs bosons mass of 300 GeV–700 GeV. Figure 51, upper, shows Feynman diagrams for the process $tbH^\pm, H^\pm \to WH, H \to \tau\tau$. Figure 51, lower left and right, shows the data, simulated background along with the signal in eighteen categories (grouped into datasets represented by the vertical dashed lines) and the limits on the $tbH^\pm$ production cross-section times $H^\pm \to WH$, $H \to \tau\tau$ branching ratio, respectively.

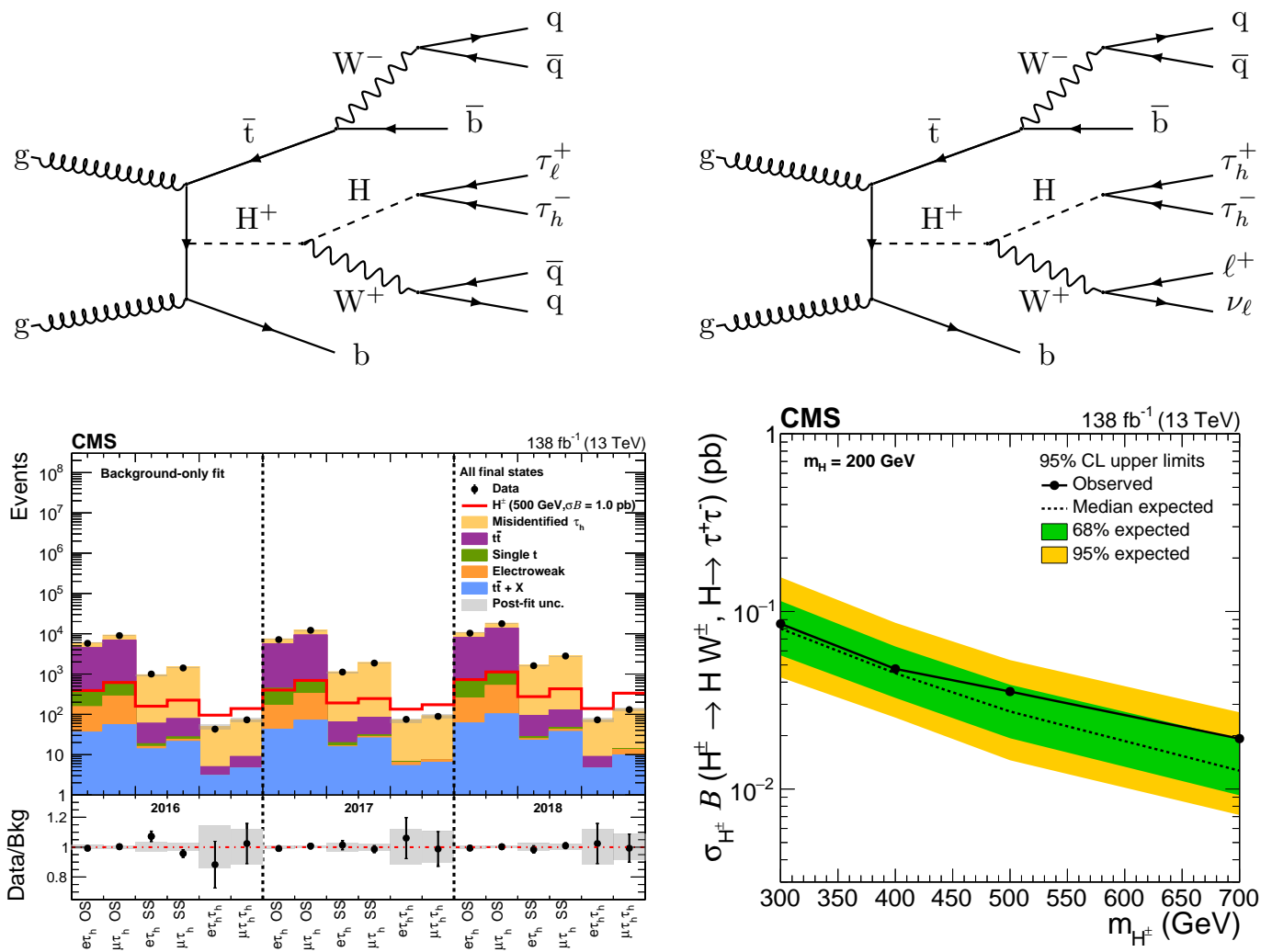

**Figure 51. Upper**: Feynman diagrams for the process $tbH^\pm$, $H^\pm \to WH$, $H \to \tau\tau$. **Lower left**: Data, simulated background and a signal in eighteen categories grouped into datasets represented by vertical dashed lines [74] (CC BY 4.0). **Lower right**: Limits on $tbH^\pm$ production cross-section times $H^\pm \to WH$ and $H \to \tau\tau$ branching ratios [74] (CC BY 4.0).

### 7.4. Double-Charged Higgs Bosons

The search for double-charged Higgs bosons, $H^{++}$, has been performed in the four-lepton final state (electrons and muons). Limits on the mass of doubly charged Higgs bosons were set at 1080 GeV at 95% CL within the left–right symmetric Type-II seesaw model [75]. Figure 52, upper left, shows a Feynman diagram of the double-charged Higgs boson boson pair production with decay into four leptons, the data, simulated background, and $H^{++}$ signals (Figure 52, lower), along with the limits on the production cross-section as a function of the $H^{++}$ mass (Figure 52, upper right).

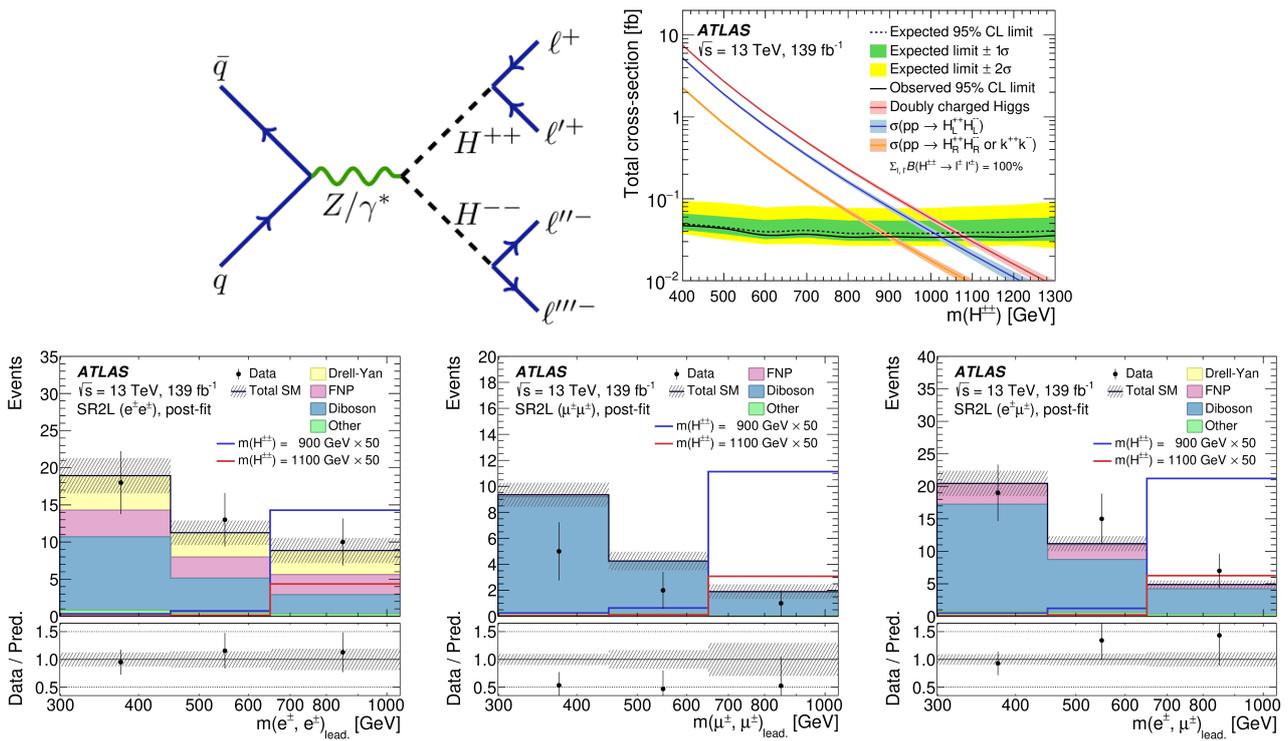

**Figure 52. Upper left**: Feynman diagram of the double-charged Higgs boson pair production with decay into four leptons. **Upper right**: Limits on the production cross-section as a function of the $H^{++}$ mass [75]. **Lower**: Data, simulated background, and $H^{++}$ signals [75] (CC BY 4.0).

## 8. Interpretations and Outlook

### 8.1. Interpretations of $H \to aa$ Searches in 2HDM+S

In the 2HDM+S Type-1 and Type-4, searches for $H \to aa$ with different final states [76] are summarized in Figure 53, left and middle.

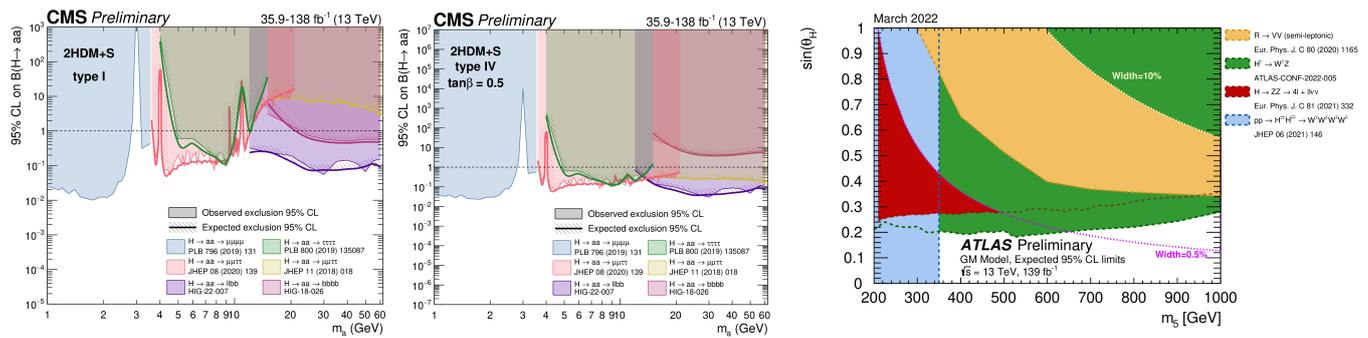

**Figure 53. Left:** Limits on the $H \to aa$ branching ratio in the 2HDM+S Typ-1 [76] (CC BY 4.0). **Middle:** Limits on the $H \to aa$ branching ratio in the 2HDM+S Type-4 [76] (CC BY 4.0). **Right:** Limits in the $(m_5, \sin \Theta_H)$ parameter plane in the Georgi–Machacek model [79] (CC BY 4.0).

### 8.2. Interpretations of Heavy Higgs Boson Searches in the Georgi–Machacek Model

Interpretations of heavy Higgs boson searches are given in the Georgi–Machacek model [77,78]. Figure 53, right, summarizes the limits in the $(m_5, \sin \Theta_H)$ parameter plane [79], thus excluding regions of the H5 plane benchmark [8].

### 8.3. hMSSM Overview

Figure 54 shows an overview of the excluded regions from the ATLAS and CMS Collaborations in the MSSM with a Higgs boson at 125 GeV [80,81].

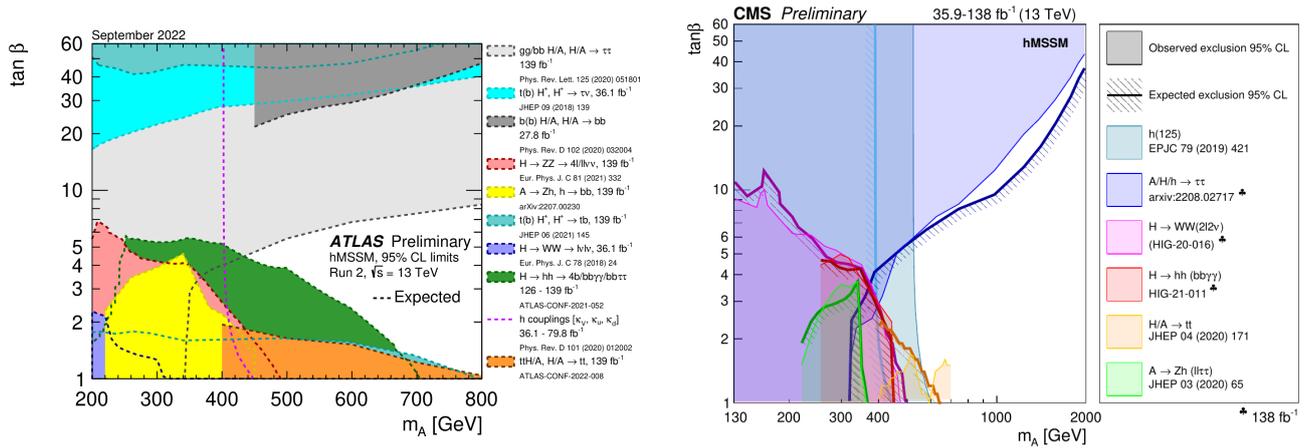

**Figure 54.** hMSSM summary of excluded $(m_A, \tan\beta)$ parameter regions by ATLAS [80] (CC BY 4.0) (**left**) and CMS [81] (CC BY 4.0) (**right**) Collaborations.

### 8.4. LHC Operation, Past, Present and Future

While the presented results were mostly based on the complete LHC Run-2 dataset, currently, LHC Run-3 operation is ongoing. A detailed overview of the LHC operations expectations from 2011 up to about 2040 [82] are given in Figure 55 [83].

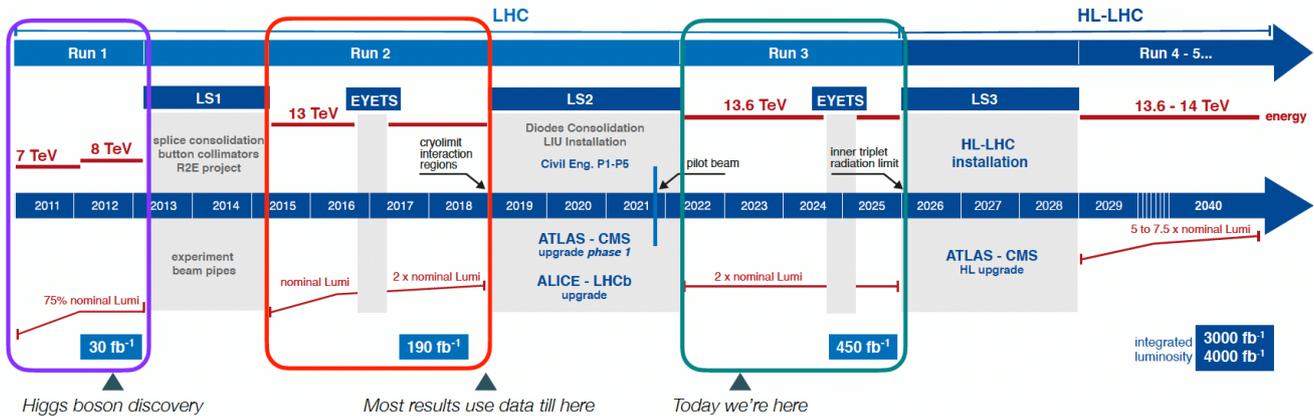

**Figure 55.** Overview of LHC operation in the past, present, and future [83] (CC BY 4.0).

### 9. Conclusions

The LHC operation has been highly successful. Over 160 fb$^{-1}$ (Run-1 and Run-2 combined) luminosity was collected by each of the LHC experiments, namely, by the ATLAS and CMS experiments. Several precision measurements of Higgs boson decay branching fractions and the Higgs boson total width were performed. The relation of Higgs boson coupling to mass is impressively linear, as predicted by the Brout–Englert–Higgs mechanism. Experimental data indicate *CP*-invariance. In searches for rare decays, LFV, dark photons, and exotics, no indication of BSM physics was observed. In modified production modes for associated production of single top and Higgs, self-coupling *HHH*, and di-Higgs production, no deviation from SM expectations were observed. Limits on new Higgs bosons and unexpected decays for additional Higgs bosons decaying into taus and photons were set, as well as for heavy Higgs bosons in *VHH* and *ttH/A* production. Furthermore, limits are set to single and doubly charged Higgs boson production. As an immediate outlook of the analysis of already recorded data, a combination of the results from the ATLAS and CMS experiments doubles statistics and thus considerably reduces the statistical uncertainties. In the current LHC Run-3 data taking period, up to 200–300 fb$^{-1}$ luminosity data is expected to be added (years 2022 to 2025). These data will be used to draw conclusions about several observed minor excesses in Higgs boson production beyond the SM. From the year 2029, HL-LHC is aiming for 3000 fb$^{-1}$, starting a new era of precision measurements. There is a highly-established and well-justified LHC program until about 2040.

**Funding:** The project is supported by the Ministry of Education, Youth, and Sports of the Czech Republic under project number LM2023040.

**Data Availability Statement:** The research data are publicly shared by the ATLAS and CMS Collaboration on https://www.hepdata.net.

**Acknowledgments:** The author would like to thank the ATLAS and CMS Collaborations for discussions and reviews of the presented results as well as the organizers of the FFK2023 Conference for their hospitality.

**Conflicts of Interest:** The author declares no conflicts of interest.

## Abbreviations

| | |
|---|---|
| ATLAS | A Toroidal LHC ApparatuS |
| BaBar | B meson and B meson antiresonance experiment |
| BDT | Boosted Decision Tree |
| BR | Branching Ratio |
| BSM | Beyond SM |
| CC BY 4.0 | Creative Commons BY 4.0 copyright |
| CERN | Conseil Européen pour la Recherche Nucléaire |
| | (European Organization for Nuclear Research) |
| CMS | Compact Muon Solenoid |
| CL | Confidence Level |
| *CP* | Charge-Parity |
| DM | Dark Matter |
| EFT | Effective Field Theory |
| exp. | expectation(s) |
| FFK | Conference on Precision Physics and Fundamental Physical Constants |
| *H* | Higgs boson |
| HDM | Higgs-Doublet Model |
| HEFT | Higgs EFT |
| HL | High-Luminosity |
| H5 | $H_5^0$ |
| hMSSM | (LHC measured Higgs mass) $m_h$ MSSM |
| ISR | Initial State Radiation |
| LEP | Large Electron-Positron (collider) |
| LHC | Large Hadron Collider |
| LHCb | LHC beauty |
| LRSM | Left-Right Symmetric Model |
| LFV | Lepton Flavor Violation |
| MET | Missing Energy Transverse |
| MSSM | Minimal Supersymmetry extension of the Standard Model |
| M125 | mass of SM Higgs boson 125 GeV |
| NMSSM | Next-to-MSSM |
| pp | proton-proton |
| QCD | Quantum Chromodynamics |
| QED | Quantum Electrodynamics |
| S | Singlet |
| SC | Signal Classification |
| SM | Standard Model |
| SS2l | Same-Sign two-lepton |
| s.d. | standard deviation(s) |
| TRSM | Two-Real-scalar-singlet extension of the SM |
| *VBF* | Vector Boson Fusion |
| WIMP | Weakly Interacting Massive Particle |
| *V* | Vector boson |
| 2/3HDM | Two/Three-HDM |